\documentclass[journal]{IEEEtran}
\usepackage{cite}

\ifCLASSINFOpdf
\else
\fi
\usepackage{amsmath}
\usepackage{amsfonts}

\usepackage{subcaption}

\usepackage{svg}

\begin{document}
%
\title{Accelerated, Memory-Efficient Far-Field Scattering Computation with
Monte Carlo SBR}
%
%
%

\author{Samuel~Audia,
  Dinesh~Manocha,~\IEEEmembership{Fellow,~IEEE},
  and~Matthias~Zwicker,~\IEEEmembership{Associate~Member,~IEEE}
  \thanks{Samuel Audia, Dinesh Manocha, and Matthias Zwicker
    are with the University of Maryland, College Park department of
    Computer Science. Dinesh Manocha is also affiliated with the
    department of Electrical and Computer Engineering. Correspondence can
be sent to sjaudia@umd.edu.}}

%
%

\markboth{Journal of \LaTeX\ Class Files,~Vol.~14, No.~8, August~2015}%
{Shell \MakeLowercase{\textit{et al.}}: Bare Demo of IEEEtran.cls for
IEEE Journals}
%



\maketitle

\begin{center}
This work has been submitted to the IEEE for possible publication. Copyright may be transferred without notice, after which this version may no longer be accessible.
\end{center}

\begin{abstract}
  We introduce a Monte Carlo integration-based Shooting and Bouncing
  Ray (SBR) algorithm for electromagnetic scattering, specifically
  targeting complex dielectric materials. Unlike traditional
  deterministic SBR methods, our approach is the first to reformulate
  the SBR integral equations using Monte Carlo techniques and
  advanced variance reduction strategies
  adapted from photorealistic rendering. This
  enables efficient, massively parallel computation on modern GPUs,
  resulting in up to a 10–15× reduction in memory usage and a 4× speed up in
  runtime, particularly for multilayer dielectric structures. Our
  method emphasizes high-energy propagation paths, efficiently
  capturing long multipath and interreflection effects.
  Verification on canonical 3D geometries and ISAR imaging of both
  conducting and dielectric representative aircraft models demonstrates that our
  Monte Carlo SBR achieves high accuracy while maintaining low noise,
  making it suitable for downstream imaging and analysis tasks.
\end{abstract}


%
\IEEEpeerreviewmaketitle

\section{Introduction}

Modern engineering designs are rooted in computer simulation to vet
and iterate performance without costly physical prototypes. In electromagnetic
scattering, simulations precisely characterize field propagation, facilitating
antenna placement \cite{nguyen_method_2013, malmstrom_effective_2017,
  whalen_antenna_2018,
johnson_genetic_1999}, RADAR analysis \cite{di_serio_2-d_2020,
liu_dynamic_2024}, and radiation safety verification
\cite{guan_electromagnetic_2017}.
Simulations, however, must balance computational speed and fidelity. Full-wave
solvers such as Method of Moments (MoM) \cite{harrington_field_1993},
Finite-Difference Time Domain (FDTD) \cite{yee_finite-difference_1997}, and
Finite Element Method (FEM) \cite{jin_finite_2008} solve the full Maxwell's
equations but require powerful computational resources and hours to weeks to
complete, often restricting their use to electrically small objects.

For objects large relative to wavelength, high-frequency approximation methods
offer a practical trade-off. Geometric Optics (GO)
\cite{pathak_electromagnetic_2022}
traces paths of plane waves and computes surface interactions semi-empirically,
but struggles with singularities at caustics and capturing wave
effects. Alternatively,
Physical Optics
(PO) \cite{pathak_electromagnetic_2022} locally approximates currents by the
immediate field incident on a planar surface. These currents enable the
use of boundary integrals, but struggle to characterize non-convex shapes and
multi-path interactions. The Shooting and Bouncing Ray (SBR) method
\cite{ling_shooting_1989, brem_shooting_2015} combines GO for incident field
propagation and PO for scattered field calculation, and has become the standard
for electrically large, complex geometries due to its speed and ease of
implementation.

SBR naturally lends itself to hardware acceleration with modern
graphics processing units (GPUs)
\cite{brem_shooting_2015, kee_efficient_2013, kim_anxel_2024}. Past
research, however, has
focused on optimizing the standard SBR algorithm through complex ray
intersection schemes
\cite{kasdorf_advancing_2021, kim_anxel_2024,yun_ray_2015} and large
tree-based ray
management \cite{brem_shooting_2015}. These approaches ignore the
underlying hardware, bottlenecking
the possible performance. GPUs prefer repeated fixed pipelines to
increase overall throughput.
Complex ray intersection need to accurately handle double counting
when ray tubes overlap adding
additional compute and communication overhead for every ray. Tracing
both the reflected and
refracted rays builds a tree recursively, making it difficult to
track the overall memory usage and
data dependencies. Furthermore, rays are traced along paths that
carry little energy and have a
negligible effect on the final result.

In contrast, photorealistic rendering
\cite{pharr_physically_2016, kajiya_rendering_1986}, for which GPUs
have been optimized,
focuses on improving Monte Carlo integration and advanced sampling techniques
\cite{veach_metropolis_1997}, increasing the overall number of rays
while decreasing their individual computation. Monte Carlo methods
have been applied to GO
ray tracing for wireless coverage mapping \cite{wang_dynamic_2022}
and sea surface
characterization \cite{peng_new_2017}. They have
not, to the authors' knowledge, been used in SBR algorithms that
incorporate physical
optics currents.

This paper introduces a novel Monte Carlo integration-based SBR algorithm for
medium to high fidelity electromagnetic scattering. Our approach departs
from traditional deterministic quadrature by randomly sampling paths
and integrating over their surface intersections, enabling efficient, trivially
parallel computation on GPU hardware. We further adapt advanced
samplings techniques
from computer graphics to a new SBR context to improve accuracy and
efficiency for
complex geometries containing dielectric materials. Unlike previous
SBR GPU implementations,
which require managing large trees of ray intersections, our method
aligns with the strengths
of GPU hardware to improve scaling of complex scenarios.

\textbf{Main Results.} Our key contributions are:
\begin{itemize}
  \item A new derivation of the SBR algorithm using Monte Carlo integration
    and a path space formalism to
    calculate the scattered field for arbitrary PEC and dielectric objects. Our
    method and deterministic SBR both converge to the same solution
    as the number
    of rays increases.
  \item Two advanced sampling techniques adapted from computer graphics
    to SBR: Fresnel-based importance sampling and Russian roulette sampling.
    We demonstrate the efficiency gains of these techniques on
    multiple mixed material objects.
  \item Implementation and benchmarking of both deterministic and Monte Carlo
    SBR on GPUs. We benchmark with 3D flat plate, sphere, dihedral, and
    airplane geometries.
    Our Monte Carlo SBR algorithm computes results within $1-2 dB$ of
    average error
    compared to deterministic SBR, while using 13 to 15 times less
    memory and 4 times faster
    runtimes on multi-layer dielectric objects.
\end{itemize}

\section{Background}

The Shooting and Bouncing Ray method and Monte Carlo integration are
well established in the literature. We provide a brief overview in
this section, but for in-depth reviews, we point readers to
\cite{brem_shooting_2015} for SBR and \cite{pharr_physically_2016}
for Monte Carlo integration in computer graphics.

\subsection{Shooting and Bouncing Ray Method}

SBR comprises three main steps: GO ray propagation of plane waves,
calculation of PO currents, and surface equivalence calculation of
scattered field. A GO wave is described by the polarization and field
intensity along a ray direction $\hat{k}$. The plane wave along the ray
is described by
\begin{equation}
  \vec{E} = \vec{E}_0e^{-jk_0\hat{k}\cdot\vec{r}},
  \label{eqn:plane-wave}
\end{equation}
where $\vec{E}^0$ is a complex vector representing the polarization and
intensity, $k_0$ is the wave
number ($\frac{2\pi f}{c_0}$) for a frequency $f$ and the speed of light
in a vacuum $c_0$. $\vec{r}$ is a field point where
the plane wave is measured. We
assume a factor of $e^{j\omega t}$ throughout. Equation
\ref{eqn:plane-wave} represents a
monochromatic wave. Rays propagate by standard GO laws using perfect
reflection and Snell's Law
\cite{born_principles_2019} for refracted rays on dielectric
boundaries. On PEC boundaries, the
internal field is assumed to be 0, so only a reflected ray is traced.
Ray intensity is attenuated
based on the Fresnel coefficient \cite{born_principles_2019} for the
corresponding polarization.

Along boundaries between homogeneous regions,
we compute the currents
using PO and the Modified Equivalent Current
Approximation (MECA) \cite{meana_modified_2010, zhang_extended_2015}.
PO makes the assumption that the current on a surface is
only a function of the incident field. In actuality, the current is a
difficult to calculate term, influenced by the radiated fields of all
other currents on the surface. Regardless, on a PEC surface, the electric PO
current is calculated by
\begin{equation}
  \vec{J} = 2\hat{n}\times\vec{H}^i,
  \label{eqn:po-pec}
\end{equation}
where $\vec{H}^i$ is the incident magnetic field and $\hat{n}$ is the
surface normal. For a plane wave,
$\vec{H}^i =  \hat{k}\times\frac{1}{\eta_0}\vec{E}^i$, where $\eta_0$ is
the free space impedance. MECA extends
this idea to dielectric surfaces using the surface equivalence
principle. On the boundary between any two homogeneous regions, the
electric, $\vec{J}$, and magnetic, $\vec{M}$, currents along the
boundary can found using
\begin{equation}
  \vec{J} = \hat{n}\times\vec{H} \; \text{ and } \; \vec{M} =
  \vec{E}\times\hat{n}.
  \label{eqn:surface-equivalence}
\end{equation}

MECA uses the relations in Equation \ref{eqn:surface-equivalence} to
approximate the electric and magnetic currents using only the
incident, reflected, and refracted fields. We denote these fields
with $i, r,$ and $t$ superscripts, respectively. Locally, the total
field is represented only by the incident GO plane wave; however, we
do not consider all three components when scattering and only sum the
fields in the region visible to the receiver. In most scattering
problems, this will be the outermost surfaces of an object. Interior
surfaces have currents associated with them, but these currents do
not need to be calculated. This leaves us with two cases to consider:
a ray entering the outer surface of an object and a ray exiting the
outer surface.

For a ray entering the outer surface, the only \textit{visible}
fields are the incident and reflected waves. MECA then approximates
the currents as
\begin{equation}
  \vec{J} = \hat{n} \times (\vec{H}^i + \vec{H}^r) \; \text{and} \;
  \vec{M} = (\vec{E}^i + \vec{E}^r) \times \hat{n}.
  \label{eqn:meca-reflect}
\end{equation}
In the exiting case, only the transmitted field is considered. For
completeness, these currents are
\begin{equation}
  \vec{J} = \hat{n} \times \vec{H}^t \; \text{and} \; \vec{M} =
  \vec{E}^t \times \hat{n}.
  \label{eqn:meca-refract}
\end{equation}
\begin{figure}[t]
  \centering
  \includegraphics[width=1.0\linewidth]{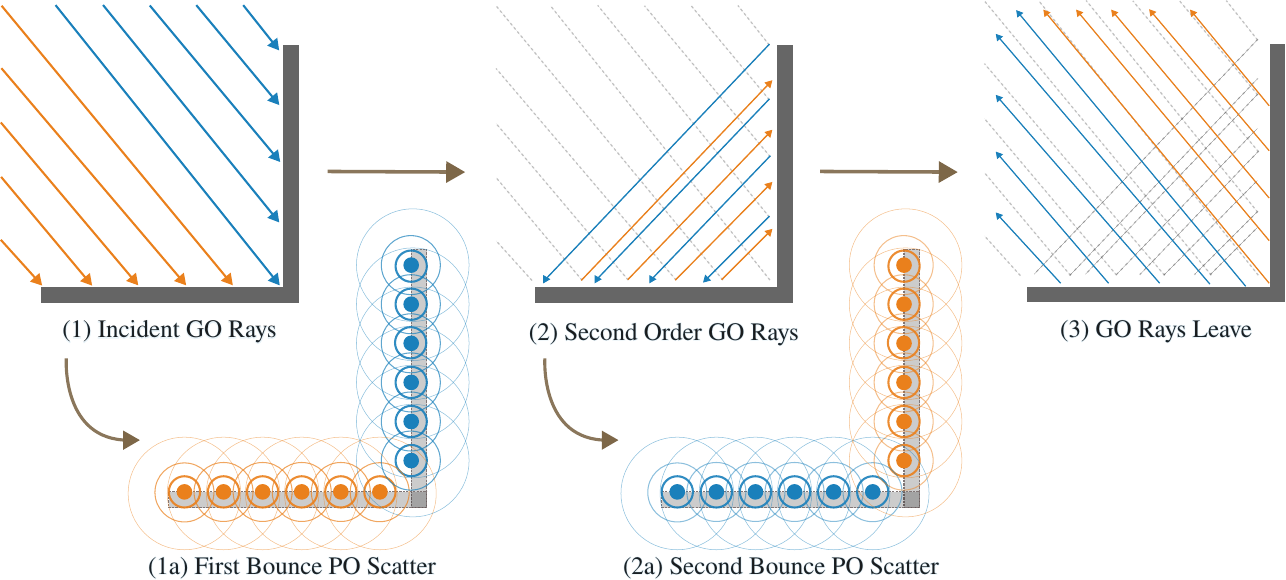}
  \caption{Core steps of the Shooting and Bouncing Ray method on a
    perfect electrical conductor dihedral. GO rays are traced
    orthographically to model a plane wave (1). At each hit, currents
    are computed and scattered to the far field using the electric
    field integral equation (1a). Rays reflect to the next surface (2),
    where currents and scattered fields are again calculated, including
    phase updates (2a). Rays exit the dihedral and no longer contribute
  to scattering (3).}
  \label{fig:core-sbr-algorithm}
\end{figure}

From the currents on the outer surface, $\Gamma^o$, the field can be computed
anywhere using the electric and magnetic field integral equation
\cite{jin_electromagnetic_2010}. Often only the far-field scattering
is desired, greatly simplifying the calculation. The far-field
electric field magnitude in the $\hat{k}^s$ direction with receiver
polarization $\hat{R}$ is calculated by
\begin{multline}
  E^s(\vec{k}^s, \hat{R}) = jk_0\eta_0\frac{e^{-jk_0r}}{4\pi
  r}\hat{R}\cdot \\
  \int_{\Gamma^o} (\hat{k}^s \times \hat{k}^s \times \vec{J} +
  \frac{1}{\eta_0}\hat{k}^s\times\vec{M})e^{jk_0\hat{k}^s\cdot\vec{r}^\prime}d\vec{r}^\prime,
  \label{eqn:scattered-e-field}
\end{multline}
where $r$ is the distance between the local phase center and the
observation point.

\subsection{Monte Carlo Integration}

Monte Carlo integration is invaluable in computing high dimensional
and difficult to discretize integrals. Instead of a Riemann integral
the equivalent expectation integral is used.
Given an arbitrary function $f(\vec{x})$,
the integral over $\Omega$ takes the form
\begin{equation}
  \int_\Omega f(\vec{x})d\vec{x} = \mathbb{E}_{\vec{x}\sim
  \mathbf{X}(\Omega)}\frac{f(\vec{x})}{p(\vec{x})}.
  \label{eqn:monte-carlo-integration}
\end{equation}

Computing the integral becomes an easier problem of drawing samples
from a distribution $\mathbf{X}$ with support on $\Omega$, denoted
$\mathbf{X}(\Omega)$, and known
probabilities $p(\vec{x})$. The only requirements are that
$\mathbf{X}$ is easily sampled and $p(\vec{x})$ is known. The
expectation can be approximated by
\begin{equation}
  \mathbb{E}_{\vec{x}\sim
  \mathbf{X}(\Omega)}\frac{f(\vec{x})}{p(\vec{x})} \approx
  (\int_\Omega d\vec{x})\frac{1}{N}\sum_{i =
  1}^N\frac{f(\vec{x}_i)}{p(\vec{x}_i)}.
  \label{eqn:mci-approximation}
\end{equation}

As $N$ tends to infinity, the approximation becomes an equality.
Convergence of the error is
estimated by the squared error, known as the variance. The variance
decreases linearly with the
number of samples \cite{pharr_physically_2016}, so the error
decreases as $O(n^{-1/2})$. Common
quadrature methods like Newton-Cotes and Gauss-Legendre, have better
convergence in lower dimensions
\cite{burden_numerical_2015}; however, as dimensionality increases,
the quadrature's time complexity
scale grows exponentially, while Monte Carlo integration error
remains independent of the
dimensionality. For SBR, Monte Carlo integration has little to no
benefit for convex PEC shapes. For
general shapes with dielectrics, the dimensionality increases with
every reflection and refraction,
tipping the scales in favor of Monte Carlo integration. We find
a runtime improvement of up to 4 times when computing nested
dielectric regions (see Table
\ref{tab:performance-comparison}).

Convergence of Monte Carlo integration can be improved by sampling
techniques like stratified sampling, metropolis sampling, or multiple
importance sampling. Additional transforms of the integral using
control variates can also improve performance. Only stratified
sampling and importance sampling is explored in this paper,
but by reformulating SBR as a Monte Carlo integration problem, we
enable future work on how to improve convergence, much like has been
seen in the computer graphics community over the last 25 years. For a
broader explanation of Monte Carlo integration in computer graphics,
we turn readers to \cite{pharr_physically_2016}.

\section{Monte Carlo Shooting and Bouncing Ray Method}

We start our derivation by considering the SBR scattering integral
in path space. From the path space perspective, it is easier to see
how each path across multiple bounces contributes to the
integral. The integral is then transformed
using Monte Carlo integration to establish our method. We
build on this novel formulation and consider stratified, Fresnel
importance, and Russian
Roulette sampling to improve performance.

\subsection{Fields in Path Space}

\begin{figure}[t]
  \begin{center}
    \includegraphics[width=0.9\linewidth]{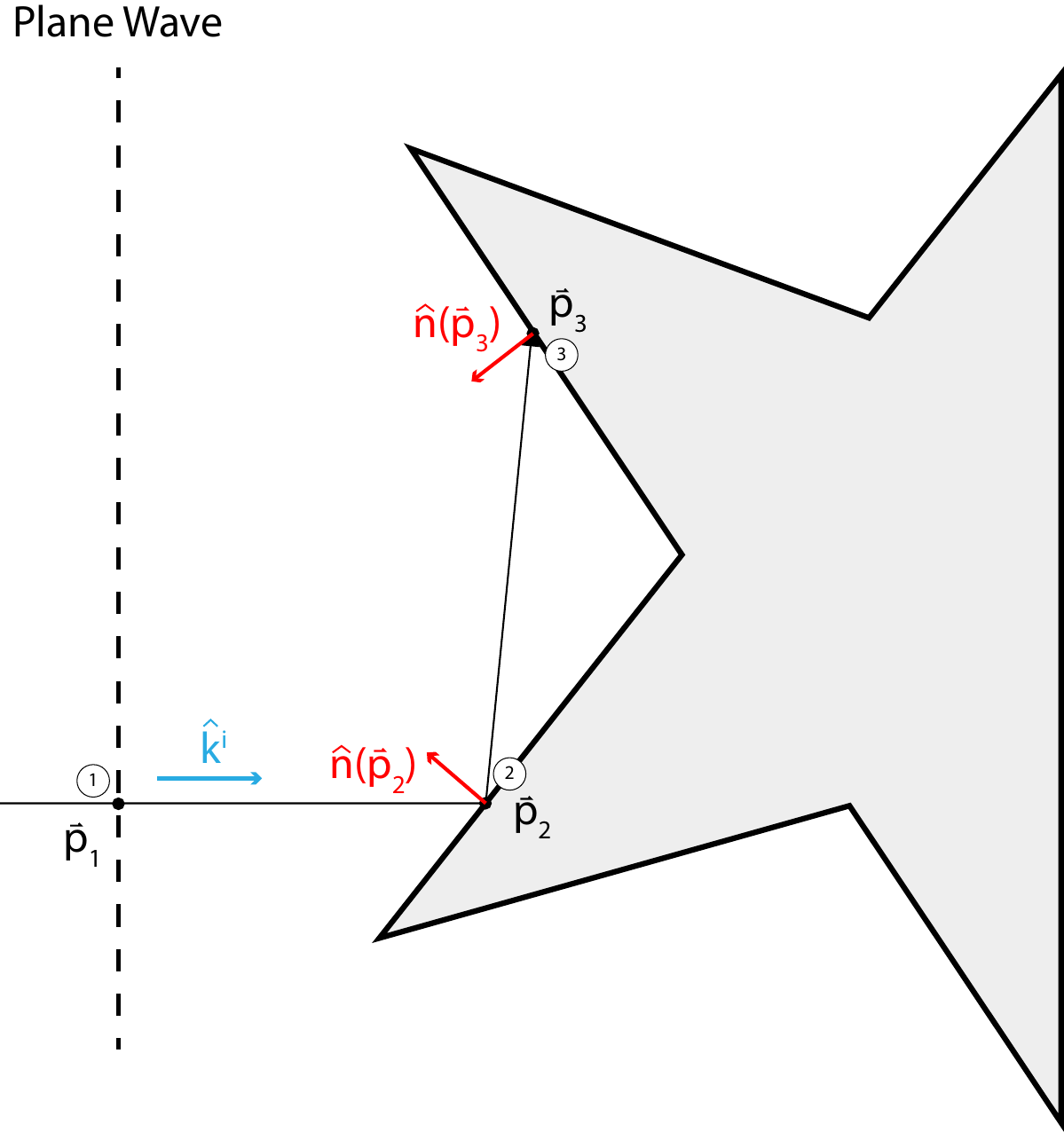}
  \end{center}
  \caption{Depiction of the path space used in our derivation. All rays start
    at the incident plane wave. Each object intersection adds a new
    dimension to the path space with corresponding normal $\hat{n}$ and
  point $\vec{r}$.}\label{fig:path-space}
\end{figure}

Consider the path space, $\mathbb{P}^N$,
instead of traditional 3D
Euclidean space, $\mathbb{R}^3$. $\vec{p}_i = (\vec{p}_{1,i}, \vec{p}_{2,i},
\ldots \vec{p}_{N,i})  \in \mathbb{P}^N$
is the $i^{\text{th}}$ path of length $N$ that connects the
transmitter and a
scattering surface current (see Figure
\ref{fig:path-space}). Each index in the path space corresponds to
an intersection with a surface, where $\vec{p}_{l,i} \in \mathbb{R}^3$ is a 3D
point on the object surface for
the $l^{\text{th}}$ intersection and $\hat{n}(\vec{p}_{l,i}) \in
\mathbb{R}^3$ is the corresponding
normal. For an incident plane wave with propagation direction
$\hat{k}^i$, we restrict all path
vectors such that $\frac{\vec{p}_2 - \vec{p}_1}{|\vec{p}_2 -
\vec{p}_1|} = \hat{k}^i$.
As we are only considering the
scattered field, we do not consider direct paths from the transmitter
to the receiver and consider
only $\mathbb{P}^N | N > 1$.

The path space is useful because we can more easily represent the
electric and magnetic fields along
each ray. We briefly drop the index $i$ for legibility. Along each
path we care about two
quantities: the phase, $\phi(\vec{p}) \in \mathbb{C}$, and the field
polarization and intensity,
$\vec{E}_0(\vec{p}) \in \mathbb{C}^3$. These quantities are defined over
the full path that the field takes. Phase is a function of the total
distance traveled and the medium
in which the field propagates. For each segment of the path, the
propagation is scaled by the
relative permittivity, $\epsilon_r$, and permeability, $\mu_r$, with
an index of refraction of $n_l
= \sqrt{\epsilon_{r,l}\mu_{r,l}}$. The overall phase is then
\begin{equation}
  \phi(\vec{p}) = \sum_{l=1}^{N-1} n_l|\vec{p}_{l+1} - \vec{p}_{l}|.
  \label{eq:phase-path}
\end{equation}

If the transmitter is in the near field, then Equation
\ref{eq:phase-path} holds; however, we want
to calculate currents from an incident plane wave, effectively
placing the transmitter at infinity,
producing an infinite phase term.  Instead, the first phase
calculation is relative to a phase
center placed at the center of the object's coordinate system. The
resulting phase is
\begin{equation}
  \phi(\vec{p}) = n_1\vec{k}^i\cdot\vec{p}_1 + \sum_{l=2}^{N-1}
  n_l|\vec{p}_{l+1} - \vec{p}_{l}|.
  \label{eq:phase-path-plane-incident}
\end{equation}

As the electric and magnetic fields propagate together in a plane wave, their
phases are the same along the total path. The field intensities and
polarizations
are also easy to track. Let $\vec{E}_0$ be the polarization
and intensity of the incident plane wave. When the field interacts
with surfaces,
the field splits, scales, and rotates based on the Fresnel equations
for the \textit{p} and
\textit{s} polarizations.  To track both
reflected and
refracted paths we introduce an additional function
\begin{multline}
  \mathbf{f}(\vec{p}_{l-1}, \vec{p}_l, \vec{p}_{l+1}) =
  \delta^r(\vec{p}_{l-1}, \vec{p}_l,
  \vec{p}_{l+1})\mathbf{T}^r(\vec{p}_{l-1}, \vec{p}_l, \vec{p}_{l+1}) + \\
  \delta^t(\vec{p}_{l-1}, \vec{p}_l,
  \vec{p}_{l+1})\mathbf{T}^t(\vec{p}_{l-1}, \vec{p}_l, \vec{p}_{l+1}),
  \label{eq:brdf}
\end{multline}
where $\delta^r$ is nonzero only when the exiting direction,
$\frac{\vec{p}_{l+1} - \vec{p}_l}{|\vec{p}_{l+1} -
\vec{p}_l|}$, is the perfect reflection of the incident
direction, $\frac{\vec{p}_l
- \vec{p}_{l-1}}{|\vec{p}_l - \vec{p}_{l-1}|}$.
Similarly, $\delta_t$ is only nonzero when the vectors satisfy Snell's law.
The 3 by 3, complex valued transformation matrices $\mathbf{T}^r$ and
$\mathbf{T}^t$ then
apply the Fresnel coefficients to the incident field. On
PEC surfaces
only the reflected term is present. Equation \ref{eq:brdf} uses a
three point form
to describe the relationship between the incident ray and exiting ray
directions by the
points that make up the two vectors. By using points instead of
normalized vector directions,
we can more easily integrate and sample surfaces. The full
expression for the GO field with respect to its path $\vec{p}$ is
\begin{multline}
  \vec{E}(\vec{p}) =
  (\prod_{l=2}^{N-1}V(\vec{p}_{l+1},
    \vec{p}_l)\mathbf{f}(\vec{p}_{l-1}, \vec{p}_l,
  \vec{p}_{l+1})) \\
  \mathbf{T}(\vec{p}_{N-1},
  \vec{p}_N)\vec{E}_0e^{-jk_0\phi(\vec{p})}.
  \label{eq:electric-plane-wave-ray}
\end{multline}
The final matrix $\mathbf{T}(\vec{p}_{N-1}, \vec{p}_N)$ is a special case. As
described in the background, we
only consider the field visible to the receiver when scattering.
Therefore, the final transformation is simply a function
of the vector from $\vec{p}_{N-1}$
to $\vec{p}_N$ and the surface normal at $\vec{p}_N$. Depending on the
orientation of the
ray and surface normal, the matrix could represent either the
reflected or refracted
field. No additional ray splitting occurs
at the final point in the path. The additional $V(\vec{p}_{l+1}, \vec{p}_l)$
term is 0 when the two points are occluded from one another and 1 otherwise.

Finally, the magnetic
field is simply the scaled field
orthogonal to the propagation direction and electric field
polarization. When considering lossy
materials, the propagation of the GO ray is slightly more complicated
\cite{brem_shooting_2015}. We
use the simpler equations in our derivation, but the extension is
straightforward.

\begin{figure}[t]
  \begin{subfigure}{0.95\linewidth}
    \begin{center}
      \includegraphics[width=1\linewidth]{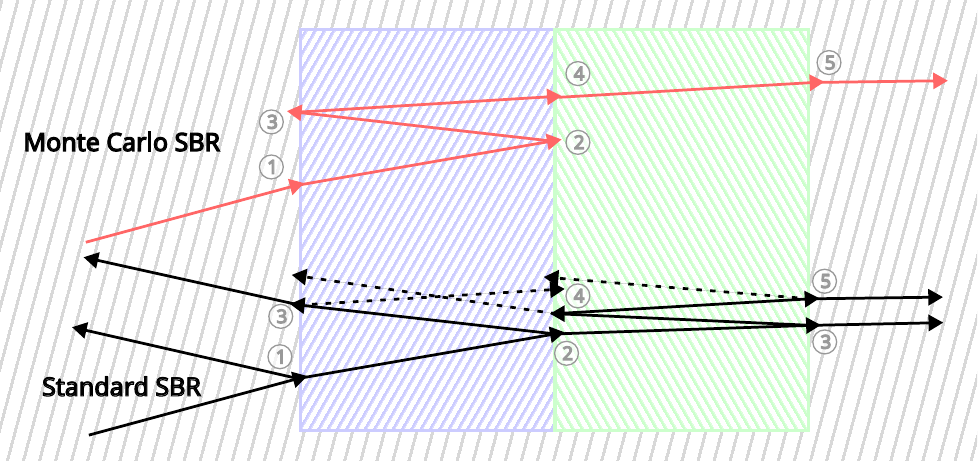}
    \end{center}
  \end{subfigure}

  \medskip
  \begin{subfigure}{0.95\linewidth}
    \begin{center}
      \includegraphics[width=1\linewidth]{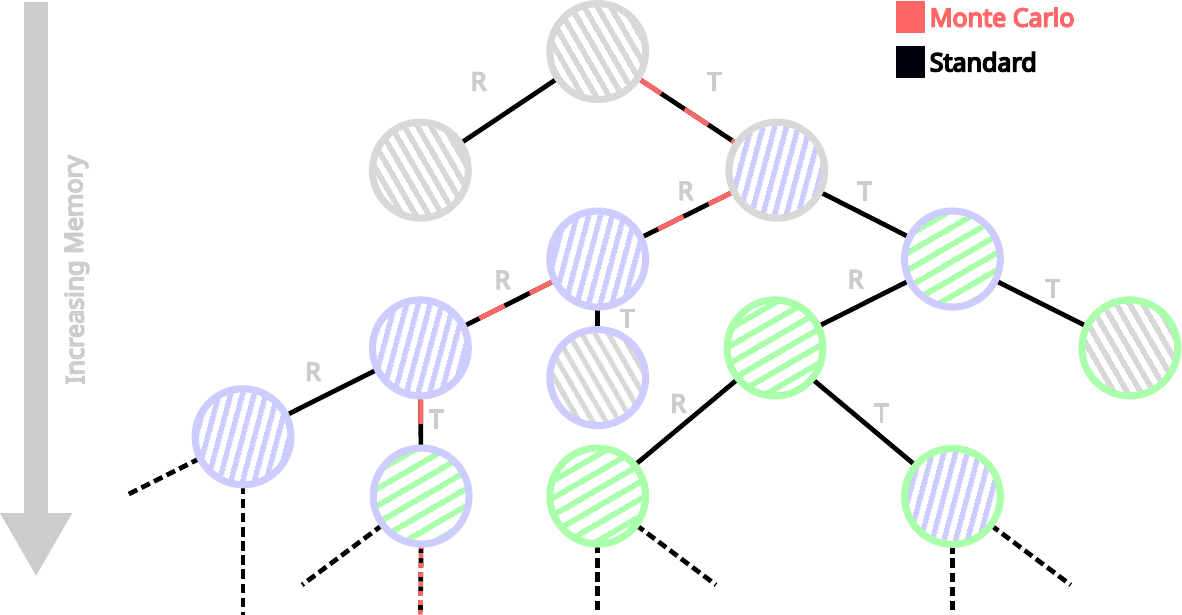}
    \end{center}
  \end{subfigure}
  \caption{Visual comparison between deterministic and Monte Carlo integration
    of dielectric regions. The deterministic algorithm must build a
    large tree to
    search the whole space, while the Monte Carlo algorithm must only
    walk a subset
  of the possible paths to calculate the scattering.}\label{fig:dielectric_tree}
\end{figure}

\subsection{Scattering Integral in Path Space}

We calculate the scattered field using Equation
\ref{eq:electric-plane-wave-ray} and Equations
\ref{eqn:surface-equivalence}, \ref{eqn:meca-reflect},
\ref{eqn:meca-refract}, and \ref{eqn:scattered-e-field}.
Far-field scattering only considers the currents on the outermost surface
which are visible to the receiver. We define a characteristic function
$\chi(\vec{p}_N)$ to only consider these currents. To clarify Equation
\ref{eqn:scattered-e-field}'s
reliance on previous path points, we express the equation as a series of nested
integrals. For all rays of path length $N$ starting
at the initial plane wave, $\Gamma_1$, the scattered field is
\begin{multline}
  E^s_N(\vec{k}^s, \hat{R}) = jk_0\eta_0\frac{e^{-jk_0r}}{4\pi r}
  \hat{R} \cdot
  \overbrace{\int_\Gamma \cdots \int_\Gamma}^{N-1} \int_{\Gamma_1}
  \chi(\vec{p}_{N}) \\
  \frac{(\hat{k}^s \times \hat{k}^s \times \vec{J}(\vec{p}) +
      \frac{1}{\eta_0}\hat{k}^s\times
  \vec{M}(\vec{p}))}
  {|\hat{k}^r_N \cdot \hat{n}(\vec{p}_N)|}
  e^{jk_0 \hat{k}^s
  \cdot \vec{p}_N} \\
  d\vec{p}_1 \cdots d\vec{p}_{N-1}dA(\vec{p}_N).
  \label{eq:scattering-continuous-path-space}
\end{multline}

After the initial plane wave, all surfaces of the object, denoted by $\Gamma$,
are integrated. The final integral basis is also changed to transform
the plane wave given by Equation \ref{eq:electric-plane-wave-ray}
in to a surface integral over the currents. This transformation introduces
a change of basis factor given by $\frac{1}{|\hat{k}^r_N \cdot
\hat{n}(\vec{p}_N)|}$,
where $\hat{k}^r_N = \frac{\vec{p}_N -
\vec{p}_{N-1}}{|\vec{p}_N - \vec{p}_{N-1}|}$. Note that the
denominator of this scale term can never be zero because that would imply
that the ray and surface are parallel. In which case, no surface intersection
occurs.

Equation \ref{eq:scattering-continuous-path-space} is written such
that intermediate path spaces are
continuous; however, SBR only traces reflected and transmitted rays
as shown in Equation \ref{eq:brdf}. As such, we
reduce the degrees of freedom
and transform the integrals into sums. These sums are over either
one or two elements depending
on if the surface is pec or dielectric, respectively. These choices
build a tree like structure of possible ray paths as visible in
Figure \ref{fig:dielectric_tree}.
There is also only a single choice for $p_2$ given $p_1$ because the incident
direction is fixed. The scattering point $p_N$ is similarly fixed by
the previous points in the path space. Each ray path, therefore, has a
maximum of $2^{N-2}$ possible configurations. The simplified scattered
field equation is then
\begin{multline}
  E^s_N(\vec{k}^s, \hat{R}) = jk_0\eta_0\frac{e^{-jk_0r}}{4\pi r}
  \hat{R} \cdot
  \overbrace{\sum^{\leq 2} \cdots \sum^{\leq 2}}^{N-2}
  \int_{\Gamma_1} \chi(\vec{p}_{N}) \\
  \frac{(\hat{k}^s \times \hat{k}^s \times \vec{J}(\vec{p}) +
      \frac{1}{\eta_0}\hat{k}^s\times
  \vec{M}(\vec{p}))}
  {|\hat{k}^r_N \cdot \hat{n}(\vec{p}_N)|}
  e^{jk_0 \hat{k}^s
  \cdot \vec{p}_N} dA(\vec{p}_1).
  \label{eq:scattering-path-space}
\end{multline}
We transform the integral basis over $p_1$ to the area
equivalent; however, no new factor is introduced because the
initial ray direction is inline with the plane wave normal by
definition. Equation \ref{eq:scattering-path-space} can be
interpreted as integrating
across all rays that reach the outer surface on their $N^{th}$ interaction.

For non-convex shapes and those containing
materials, a path may intersect the outer surface multiple times. As such,
we must consider paths of all lengths and the full scattered field is
given by
\begin{equation}
  E^s(\vec{k}^s, \hat{R}) = \sum_{l=2}^\infty E^s_l(\vec{k}^s, \hat{R}).
  \label{eq:scattered-field-sum}
\end{equation}

The combination of Equations \ref{eq:scattering-path-space} and
\ref{eq:scattered-field-sum}
fully characterizes the scattered field computed by the SBR algorithm.
The novel path space interpretation of the SBR integral highlights
the difficult scaling of the algorithm. As longer and longer path
lengths are considered, $2^{N-2}$ grows exponentially. We depict this
growth for two dielectric regions sharing a single boundary in Figure
\ref{fig:dielectric_tree}.
The deterministic
SBR algorithm must trace all of these growing number of paths, while
our Monte Carlo method only explores a single path.

\subsection{Monte Carlo Integration of Scattering}

Equation \ref{eq:scattered-field-sum} can be computed either using Riemann
integration \cite{brem_shooting_2015} or with our method using Monte Carlo
integration. Previous works required the tracing of all paths. This requires
difficult tracking of beam areas and complicated field discretization
schemes \cite{kasdorf_advancing_2021, kim_anxel_2024}.
Instead, our method simplifies the computation for each ray and does not
require tracking multiple inter-reflections in dielectric mediums. Our method
can result in more overall rays being traced; however, the lack of
stack is easier to parallelize on GPU hardware, often resulting in
better performance.

Applying Monte Carlo integration to Equation
\ref{eq:scattering-path-space}
results in
\begin{multline}
  E^s_N(\vec{k}^s, \hat{R}) = jk_0\eta_0\frac{e^{-jk_0r}}{4\pi r}
  \hat{R} \cdot \\
  \mathbb{E}_{\vec{p} \sim \mathbb{P}^N}
  \frac{\chi(\vec{p}_{N})}{p(\vec{p})} \frac{(\hat{k}^s \times \hat{k}^s
      \times \vec{J}(\vec{p}) +
  \frac{1}{\eta_0}\hat{k}^s\times\vec{M}(\vec{p}))}{|\vec{k}^r_N\cdot\hat{n}(\vec{p}_N)|}
  \\
  e^{jk_0 \hat{k}^s
  \cdot \vec{p}_N},
  \label{eq:scattering-path-space-monte-carlo}
\end{multline}
where $p(\vec{p})$ is the probability of the sampled path.
Herein lies the benefit of our method. If we are able to sample,
with a known probability, rays that contribute the most to the scattered field,
we can improve the performance of the SBR algorithm. The probability of the
full path is the product of the probability at each path segment, giving
\begin{equation}
  p(\vec{p}) = \prod_{l=1}^N p(\vec{p}_l).
  \label{eq:path-probability}
\end{equation}

For PEC reflections, there is only one possible
path the ray can take with a probability of 1.
Equation \ref{eq:scattering-path-space-monte-carlo} is then discretized
for numerical integration, giving
\begin{multline}
  E^s_N(\vec{k}^s, \hat{R}) \approx jk_0\eta_0\frac{e^{-jk_0r}}{4\pi r}
  \frac{\int_{\Gamma_1} dA(\vec{p}_1)}{N_{s}} \hat{R} \cdot \\
  \sum_{i=1}^{N_{s}}
  \frac{\chi(\vec{p}_{N,i})}{p(\vec{p}_i)} \frac{(\hat{k}^s \times \hat{k}^s
      \times \vec{J}(\vec{p}_i) +
  \frac{1}{\eta_0}\hat{k}^s\times\vec{M}(\vec{p}_i))}{|\vec{k}^r_{N,i}\cdot\hat{n}(\vec{p}_{N,i})|}
  \\
  e^{jk_0 \hat{k}^s
  \cdot \vec{p}_{N,i}},
  \label{eq:discretized-path-space-mci}
\end{multline}
where $N_{s}$ is the number of paths sampled and $\int_{\Gamma_1} dA(\vec{p}_1)$
is the sampled cross-sectional area of the incident plane wave.
To compute each term of the infinite sum in Equation
\ref{eq:scattered-field-sum},
the path from the previous term is extended by tracing another ray. This
approach removes the need to recalculate existing paths without
biasing the estimator.
Two factors
contribute the most to the path
in SBR: the initial ray launch location and dielectric interactions.
In the following subsections,
we discuss how to compute the initial distribution and advanced
sampling methods to improve
performance of our method.

\subsection{Stratified Sampling}

\begin{figure}[t]
  \begin{center}
    \includegraphics[width=0.95\linewidth]{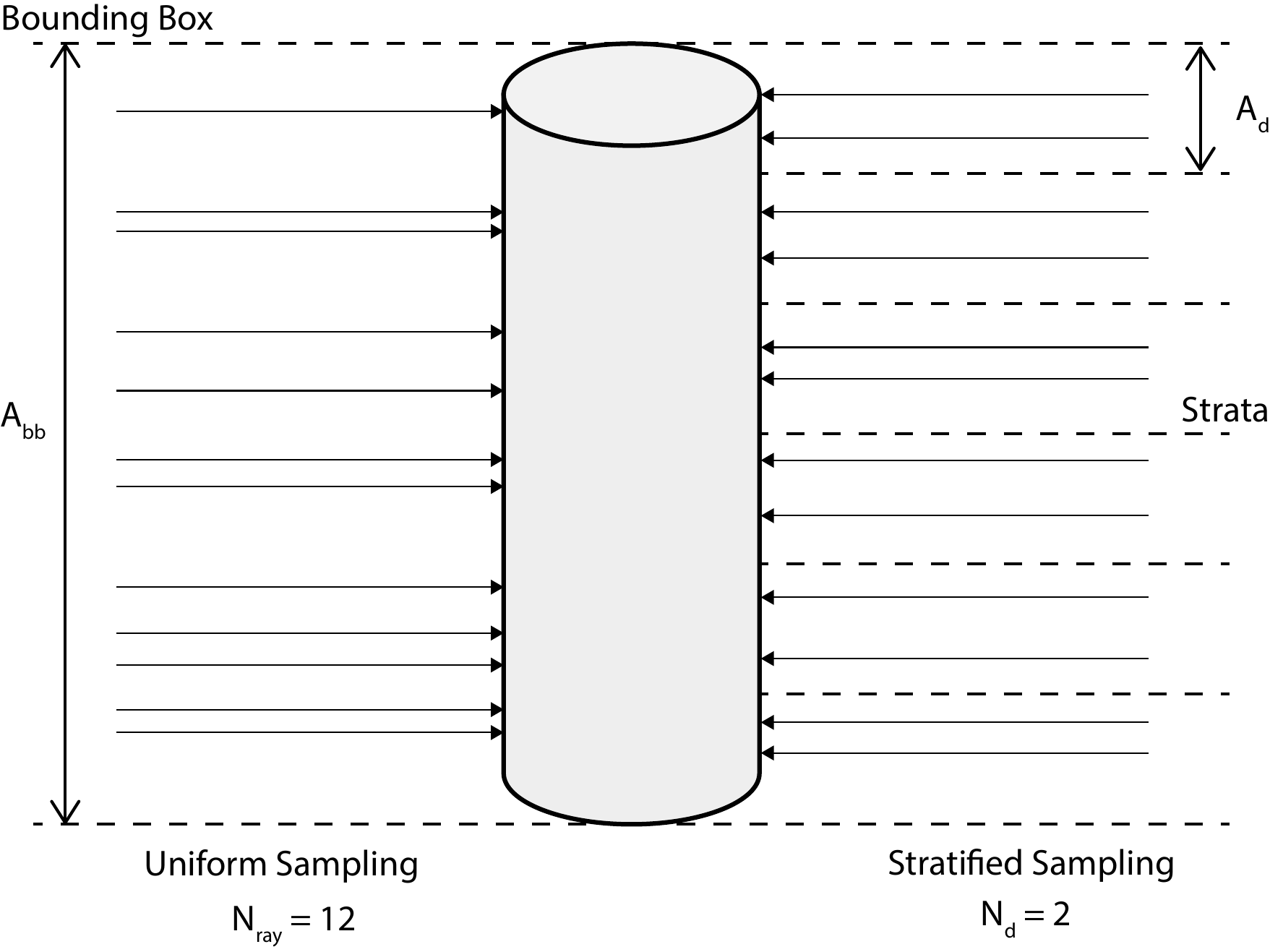}
  \end{center}
  \caption{Comparison between uniform (left) and stratified (right)
    sampling. Uniform
    sampling can produce unexpected cluster of samples. Stratified
    sampling mitigates
    the issue by uniformly sampling in subdivisions called strata. In
    SBR this allows
    the algorithm to more easily sample the oscillations of induced
  currents.}\label{fig:stratified-vs-uniform}
\end{figure}

For the initial ray launch, a bounding
box is calculated around the
object by projecting the computational mesh on the plane orthogonal
to the incident ray. Ray
starting positions are then sampled from this 2D rectangle. If
sampled uniformly, the probability of
each ray is simply one over the area of the bounding rectangle,
$A_{bb}$.
In our implementation, we found
that simple uniform sampling produced results that were too noisy
compared to the deterministic
algorithm. To mitigate the problem, we apply stratified sampling
\cite{pharr_physically_2016}.

Instead of every ray sampling from the whole bounding box, the
box is first subdivided into strata. Each stratum is sampled
$N_d$ times and has an area of $A_d = \frac{A_{bb}}{N_{s} / N_d}$
(Figure \ref{fig:stratified-vs-uniform}).
As an analog to computer graphics, imagine the plane wave is discretized
into pixels. Each pixel is sampled independently. The size of these
individual strata along with their sampling is explored in Subsection
\ref{sec:convergence}. Finer sampling improves
integration of the surface current oscillations, and we find that
strata on the order of a wavelength perform best.

\subsection{Fresnel Importance Sampling}

Unlike in deterministic SBR, Monte Carlo SBR picks a single reflected
or refracted ray at each
dielectric boundary instead of tracing both. This removes the need to
build a tree of paths,
simplifying the algorithm and making SBR trivially parallelizable as
all path contributions are
independent without needing to backtrack. Monte Carlo implementations
have flexibility in how they
choose the reflected or refracted path. A simple option is give each
a direction a probability of
0.5; however, this ignores the relative energy along the two paths.
For example, a ray entering
glass from air at normal incidence will reflect with a twentieth of
the power. The rest in
transmitted into the medium. Intuitively we want to sample along rays
which carry the most energy.
Doing so is easily accomplished by adjusting the probabilities at the
dielectric boundaries based on
the computed Fresnel coefficients. For our
implementation, we set the
probability of a reflected ray to the averaged Fresnel coefficient,
$\bar{R}$, for each polarization
giving
\begin{equation}
  \bar{R} = \frac{|r_s| + |r_p|}{2}.
  \label{eq:average-fresnel-sampling}
\end{equation}

The probability of refracted rays is $1 - \bar{R}$. There are
two important characteristics of this probability. First, we
purposely chose to use Fresnel coefficient directly instead of ray
power because the square in the power often oversamples the reflected
or refracted path. For example, in the air to glass example above, the power is
0.04 while $\bar{R}$ is 0.2. Second, the reflection coefficients
must be used because they are bounded between 0 and 1. Transmission
coefficients can be greater than 1 when transitioning into a
medium with a lower index of refraction. Using reflection coefficients
also more easily aligns with PEC surfaces where the reflection
coefficient is always 1. The effect of this sampling technique
is explored in Section \ref{sec:fresnel-sampling}.

\subsection{Russian Roulette Sampling}

Equation \ref{eq:scattered-field-sum} shows that we must consider
path lengths up to infinity. If we
simply terminated the ray at a set number of bounces, then the Monte
Carlo integration would be
biased and converge to a different integral value. In practice,
however, we will need to set a
maximum bounce limit. Ideally we can set this limit to a very high
value to reduce the error, but at
the same time, we do not want to trace long paths that have little
energy, increasing runtime. We
use Russian roulette sampling to kill off rays with little energy and
average their contribution
into the other rays. Before adding another segment to each path, a
probability distribution is
randomly sampled to determine whether to continue. This decision is
determined by a threshold
$q \in (0, 1)$. By convention, if the drawn value is greater than
$q$, then the path
continues. To not introduce bias into the integral, all rays that continue
are weight by $\frac{1}{1-q}$. In our work, we set $q = 0.5$ and draw
from a uniform
distribution. Therefore, continuing each path is, in effect, a coin
flip. Future work
could explore different cutoffs and sampling distributions. The
effect of Russian Roulette sampling is explored in
Section \ref{sec:russian-roulette-sampling}.

\subsection{Hardware Alignment}

For GPU parallelization, SBR algorithm has
two main alternating stages (Figure \ref{fig:core-sbr-algorithm}):
the trivially parallel ray trace to compute PO currents from the GO
field and the reduction to
calculate the scattered
field. Ray-surface intersections have
been extensively
studied and optimized. Generally, an acceleration structure like a k-d tree
\cite{bentley_multidimensional_1975} or bounded volume hierarchy (BVH)
\cite{meister_survey_2021} is used to reduce the number of triangles that need
to be queried. From there, a simple 3 × 3 linear equation is solved for each
triangle to check for ray intersection. This traversal and intersection
calculation are what the application specific integrated circuits (ASIC) ray
tracing cores compute. General computation, such as computing currents or new
ray directions, are executed on the general GPU compute cores with no ASIC
improvements. There is little difference in the these steps between the Monte
Carlo and standard SBR algorithms. What does change is removal of a call stack
for dielectric paths. The standard algorithm traces all reflected
and refracted
paths until the path energy drops below a certain threshold. This traversal is
similar to a depth first search and requires the construction of a
stack to save
the field’s state at each junction. The stack adds additional memory
overhead to
the ray trace, limiting the length of paths. The varying execution paths
also cause divergence between GPU threads, lowering throughput as
groups of threads must execute
the same instruction. If too many adjacent operations differ, then GPU threads
can be left inactive waiting their turn. The
Monte Carlo
integration algorithm, in contrast, does not require saving the field state at
junctions because it simply picks one of the paths and proceeds.
This simplicity
makes it easy to implement an iterative ray tracer and enable greater
throughput.

The reduction step can be difficult to parallelize on the GPU as it requires
communication between threads. The recursive nature of the
deterministic algorithm
makes it difficult to predetermine the data access patterns of the
algorithm. The
Monte Carlo algorithm, in contrast, uses a fixed number of paths per GPU call.
As such, the communication and memory write overhead can be better amortized
by coalescing nearby loads and stores. Overall, the Monte Carlo algorithm
is easier to vectorize than the deterministic algorithm by advancing
a wavefront of rays without needing to backtrack.

\begin{figure}[t]
  \begin{subfigure}{1.0\linewidth}
    \centering
    \includegraphics[width=0.8\linewidth]{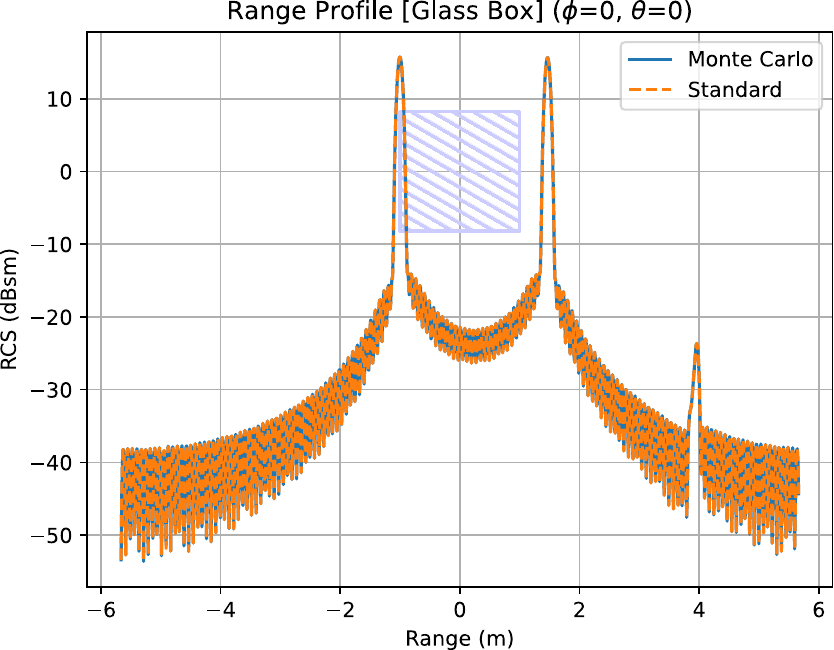}
  \end{subfigure}

  \medskip
  \begin{subfigure}{1.0\linewidth}
    \centering
    \includegraphics[width=0.8\linewidth]{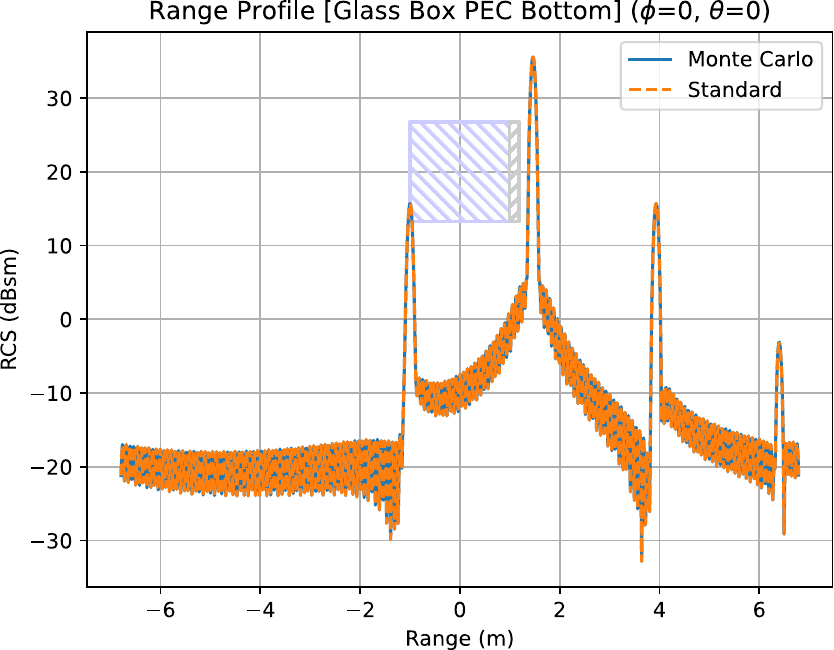} \end{subfigure}
  \caption{Range profile correctness for dielectric cubes with and
    without PEC backing. The peaks match between deterministic and Monte
    Carlo SBR on multiple interreflections and scattering of the
  field.}\label{fig:correctness-dielectric}
\end{figure}

\begin{figure}[thp]
  \centering
  \begin{subfigure}{1.0\linewidth}
    \centering
    \includegraphics[width=0.8\linewidth]{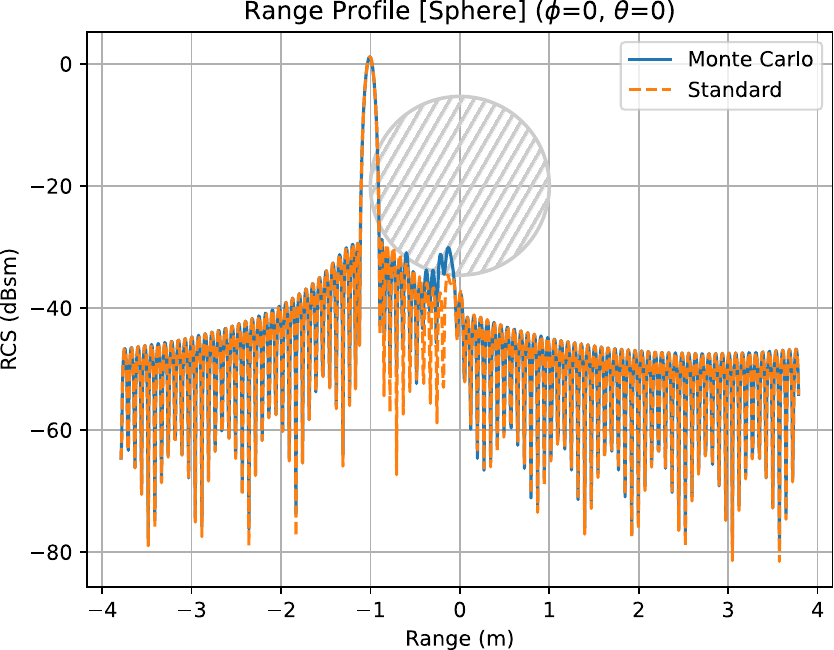}
  \end{subfigure}

  \smallskip
  \begin{subfigure}{1.0\linewidth}
    \centering
    \includegraphics[width=0.8\linewidth]{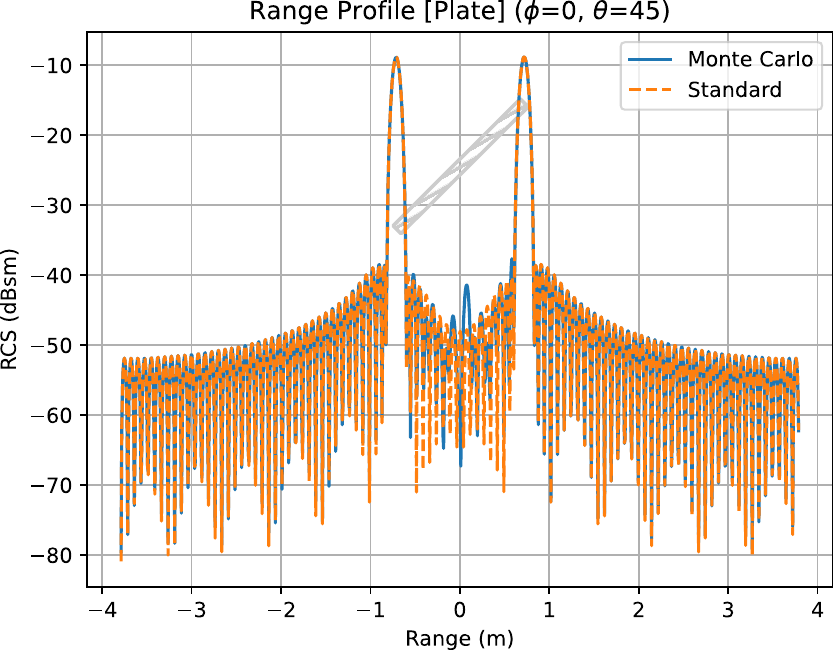}
  \end{subfigure}

  \smallskip
  \begin{subfigure}{1.0\linewidth}
    \centering
    \includegraphics[width=0.8\linewidth]{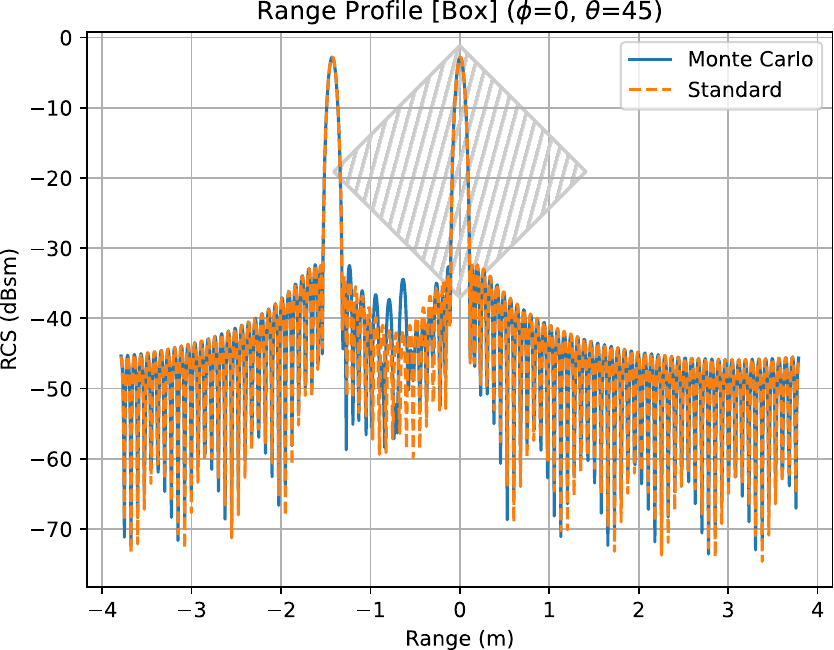}
  \end{subfigure}

  \smallskip
  \begin{subfigure}{1.0\linewidth}
    \centering
    \includegraphics[width=0.8\linewidth]{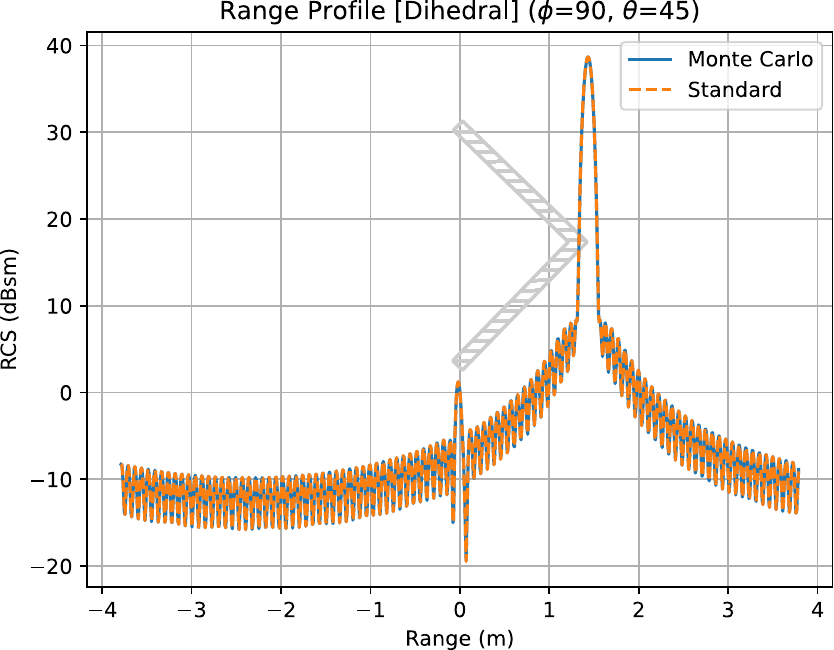}
  \end{subfigure}
  \caption{These range profiles compare the correctness of the
    deterministic and Monte Carlo SBR
    algorithms. We see that peaks align at the expected range
    location with little noise introduced by the random
  sampling.}\label{fig:range-profile-mci-vs-std}
\end{figure}

\begin{figure}[t]
  \begin{subfigure}{1.0\linewidth}
    \centering
    \includegraphics[width=0.8\linewidth]{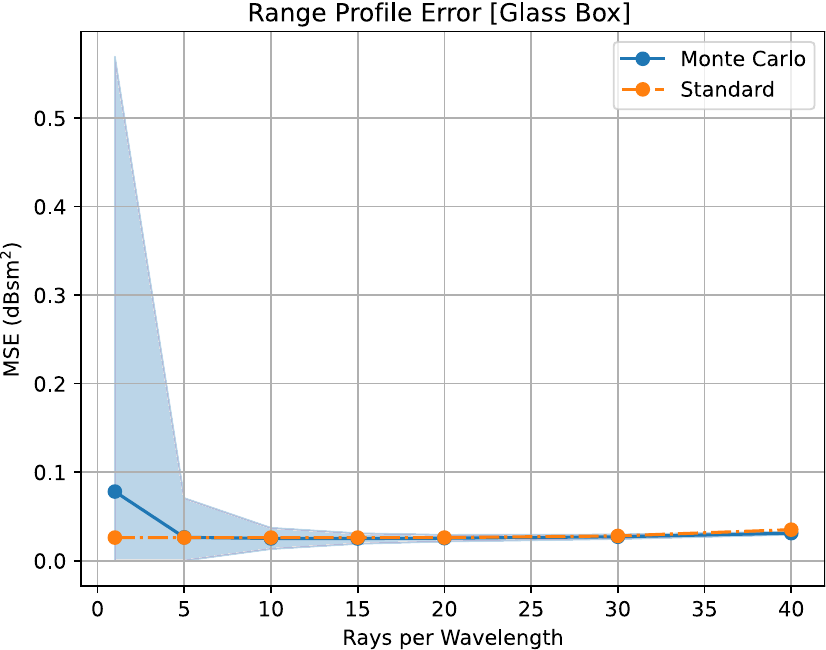}
  \end{subfigure}

  \medskip
  \begin{subfigure}{1.0\linewidth}
    \centering
    \includegraphics[width=0.8\linewidth]{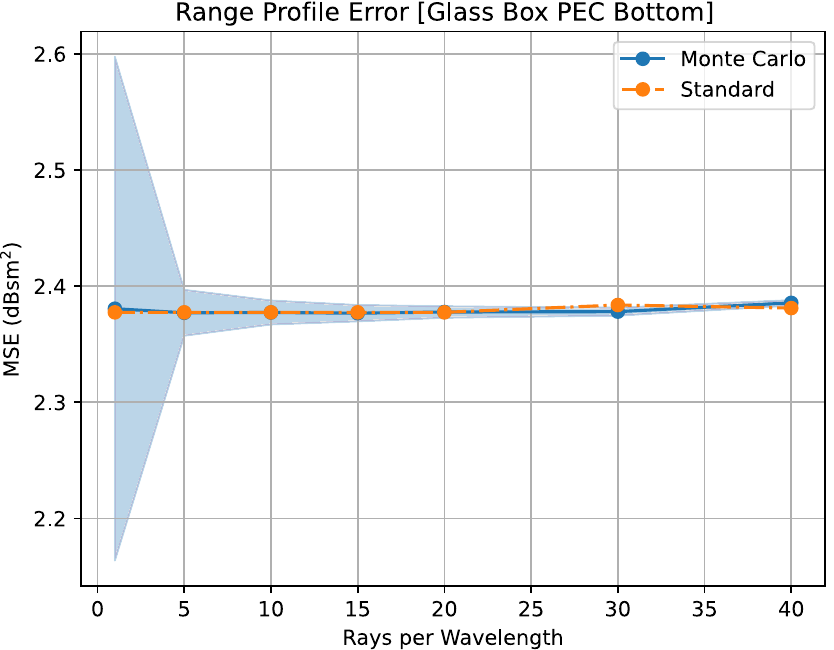}
  \end{subfigure}
  \caption{Error in dB space across 100 runs of the Monte Carlo
    algorithm as a function
    of the average number of rays per wavelength. As expected, as the
    number of rays
    increases, the deterministic and Monte Carlo algorithms converge
    to the same answer.
  Ground truth is given by the analytical solution.}
  \label{fig:std-mci-convergence-comparison-exact}
\end{figure}

\section{Results}

We evaluate the performance of our Monte Carlo SBR algorithm against
deterministic SBR on a variety of 3D benchmarks including: a flat plate,
a sphere, a box, a dihedral, a nested dielectric, and an airplane. We
also investigate the effect of Fresnel splitting and
Russian roulette sampling on dielectric materials. Suitability for
downstream signal processing is demonstrated with inverse synthetic
aperture RADAR (ISAR) images of a PEC and dielectric airplane
geometry.

We use the Mitsuba \cite{jakob_mitsuba_2022} retargetable
renderer to implement both methods. Mitsuba has been used extensively
in computer graphics research \cite{zeltner_specular_2020,
guillen_general_2020, nicolet_recursive_2023}, and provides Fresnel
equations and ray triangle intersections. The renderer is build on
top of Dr. Jit \cite{jakob_drjit_2022} which uses just-in-time
compilation to optimize the solver and execute on GPU hardware using
CUDA \cite{nickolls_scalable_2008} and OptiX
\cite{parker_optix_2010, kee_efficient_2013}.
Our implementation is not simply a direct application of Mitsuba, which
does not currently support surface current calculation or physical
optics integration. We rather, implemented our own render and utilized
the lower level functionality such as GPU dispatch and random number
generation. All experiments are run using an RTX 2070
super GPU with 8 GB of video memory.

\subsection{Correctness}

\begin{figure}[thp]
  \begin{subfigure}{1.0\linewidth}
    \centering
    \includegraphics[width=0.8\linewidth]{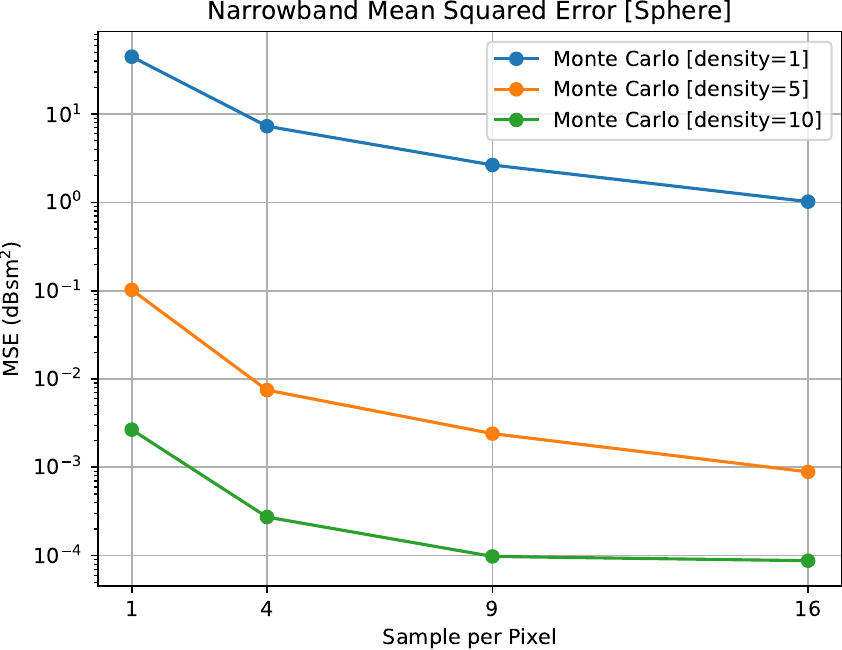}
  \end{subfigure}

  \smallskip
  \begin{subfigure}{1.0\linewidth}
    \centering
    \includegraphics[width=0.8\linewidth]{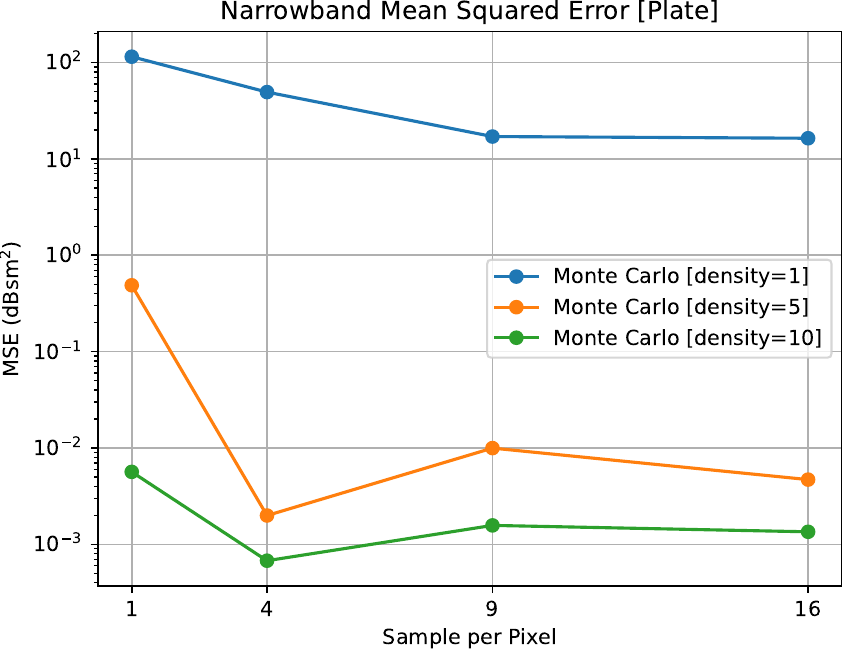}
  \end{subfigure}

  \smallskip
  \begin{subfigure}{1.0\linewidth}
    \centering
    \includegraphics[width=0.8\linewidth]{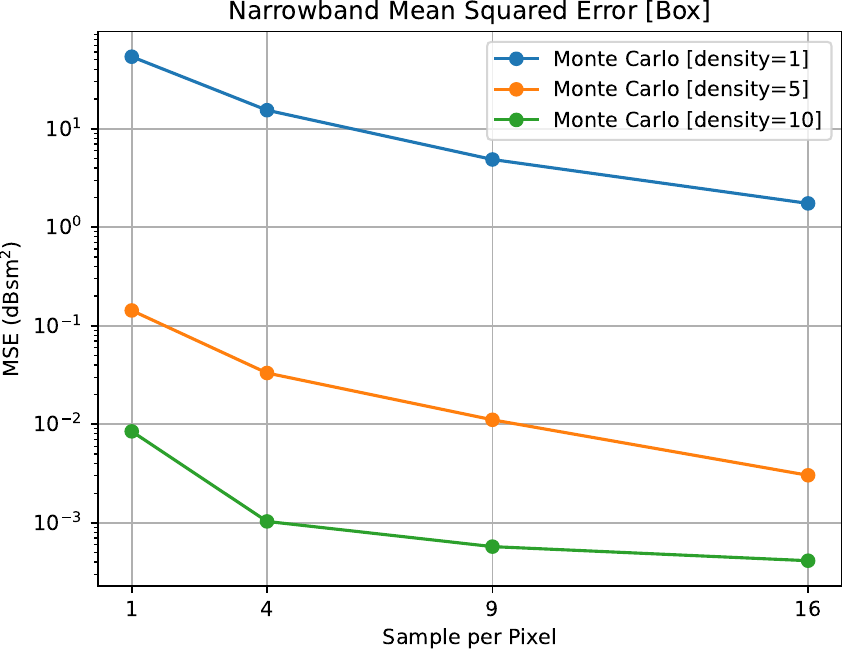}
  \end{subfigure}

  \smallskip
  \begin{subfigure}{1.0\linewidth}
    \centering
    \includegraphics[width=0.8\linewidth]{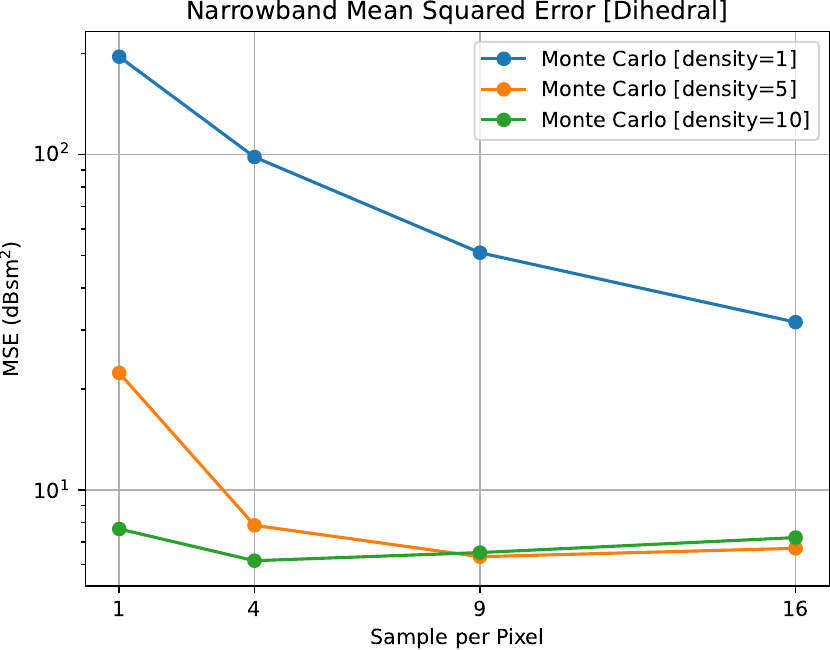}
  \end{subfigure}
  \caption{Error convergence versus sampling density for VV
    narrowband sweeps of various geometries. Strata area strongly
    influences performance, with in-beam sampling providing additional
  refinement.}
  \label{fig:sampling-convergence}
\end{figure}

We begin by showing that the deterministic SBR algorithm and our
Monte Carlo SBR
algorithm converge to the same results and produce the same range
profile for PEC
and dielectric objects. Range profiles are constructed by taking the
inverse-fast-Fourier transform of the frequency data and scaling by
the speed of
light in a vacuum. The square root radar cross section (RCS) of the field,
$\sqrt{\sigma} = \lim_{r\rightarrow\infty}
2\sqrt{\pi}r\frac{\vec{E}^s}{|\vec{E^{i}}|}$,
is used to push the receiver location to infinity while maintaining complex
phase information. Range profiles are highly dependent on the
continuity, amplitude, and phase
of the computed field across multiple frequencies demonstrating our
methods suitability for
common radar processing algorithms. Additionally, range profiles
provide easy to visualize
comparisons with clear connections with the underlying geometry.
Peaks correspond to scattering features like current
discontinuities and specular reflections.

Frequencies between 1 and 3 GHz were
computed with a 20 MHz step size. The sphere, plate, cube, and dihedral are
modelled as PEC. The glass cubes are represented as bulk materials with
$\epsilon_r=1.5$. We set the ray density to $\frac{\lambda}{40}$ in
the deterministic
algorithm and stratified beams to $\frac{\lambda}{10}$ with 16 samples in each
stratum. The VV response is considered in all plots,
corresponding to the polarization of the receiver and transmitter,
respectively. Both algorithms are using the same overall number of
rays. We use a
simple midpoint rule for integrating over the beam area in the classic
algorithm. Our method compares favorably to the standard algorithm
(Figure \ref{fig:correctness-dielectric} and
\ref{fig:range-profile-mci-vs-std}).

For the
dihedral, the glass cube, and the PEC backed glass cube, the two
techniques are
line for line. The sphere, plate, and PEC cube, match at the peaks, but there is
some additional noise in the Monte Carlo solution. These regions correspond to
parts of the object that need to cancel adjacent sources in phase. This
additional noise is the main drawback of the Monte Carlo integration method;
however, the noise can be easily lowered by taking more samples. We
also do not
believe that this noise will hinder downstream processing.

\subsection{Convergence}
\label{sec:convergence}

Convergence is explored by computing the mean squared error of
narrowband plots for
the sphere, plate, PEC box, and dihedral swept from 0 to 90 degrees
in $\theta$. The error is computed relative to the deterministic method solved
with a density of $\frac{\lambda}{60}$. We compute the VV
response at 101 frequencies from 1 to 3 GHz. The equivalent deterministic
algorithm ray density can be computed as $density \cdot \sqrt{samples
\; per \; pixel}$. Figure
\ref{fig:sampling-convergence} shows that by increasing the strata ray
density, the error is
greatly reduced, while increasing the number of samples taken within
each sub beam decreases error
by a smaller magnitude. From these results, we recommend sizing the
strata areas to 10 rays per
wavelength and draw multiple samples within these sub beams as needed.

For the dielectric cubes, we are able to
compare range profiles to the analytical solutions at normal incidence
by matching the phases at each boundary and computing
the scattering. Figure \ref{fig:std-mci-convergence-comparison-exact}
shows the prediction error of the standard and Monte Carlo algorithms as
compared to the exact solution. To characterize
the Monte Carlo estimate, we average
the error across 100 initial seed values and plot the mean and first
standard deviation. When sampling with very low ray densities, the
variability of the Monte Carlo estimate is high with swings of up to
$0.5 \; dBsm^2$; however, once at least 10 rays per wavelength are
sampled, the variance is greatly reduced, and the two algorithms are
within $0.01 \; dBsm^2$ of error. We have demonstrated that our method is
able to converge to a reasonable answer using a comparable number of
rays as the standard algorithm.

\subsection{Runtime and Memory Usage}

\begin{table}
  \centering
  \caption{Runtime and memory comparison of Monte Carlo and standard
    SBR methods on various dielectric shapes, using 20 samples per
    wavelength. Runtimes are averaged over 10 runs (plus 3 warmup runs
    for GPU kernel caching). All dielectric shapes benefit from
    lower memory usage, while nested dielectrics show improved runtime.
    Dr. Jit employs a copy-on-write optimization, given in
    parentheses, which is challenging to replicate in lower-level
    code. SBR memory usage is characterized by forcing allocation at
  each material boundary and reporting both values.}
  \label{tab:performance-comparison}
  \begin{tabular}{c|cc|cc}
    & \multicolumn{2}{c|}{Monte Carlo (ours)} &
    \multicolumn{2}{c}{Deterministic} \\
    Units & s $\downarrow$ & MiB $\downarrow$ & s $\downarrow$& MiB
    $\downarrow$\\
    \hline
    Glass Cube & $\mathbf{2.47 \pm 0.01}$ & \textbf{10.8} & $2.70 \pm 0.08$ &
    128.3 (19.3) \\
    PEC Bottom & $2.78 \pm 0.02$ & \textbf{12.0} & $\mathbf{2.52 \pm 0.08}$ &
    128.3 (19.3) \\
    Nested & $\mathbf{2.82 \pm 0.02}$ & \textbf{24.2} & $11.0 \pm
    0.10$ & 348.2 (88.8)\\
  \end{tabular}
\end{table}

\begin{figure}[t]
  \centering
  \begin{subfigure}{0.49\linewidth}
    \centering
    \includegraphics[width=0.9\linewidth]{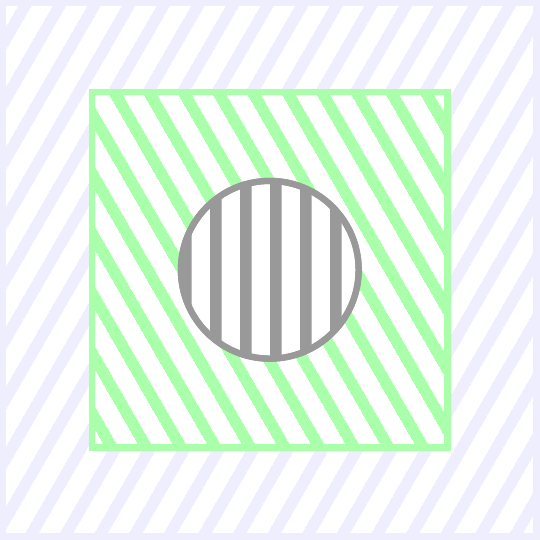}
  \end{subfigure}
  \begin{subfigure}{0.49\linewidth}
    \centering
    \includegraphics[width=0.9\linewidth]{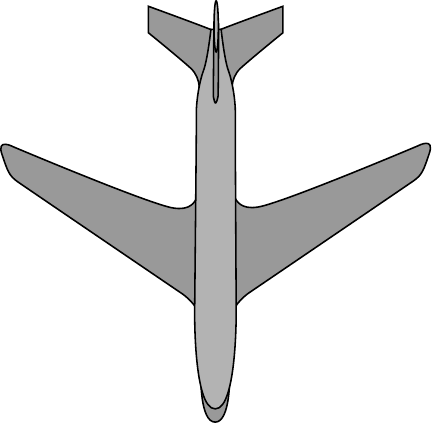}
  \end{subfigure}
  \caption{Illustration of test geometries: Left, a layered
    dielectric cube ($3\,$m, $\epsilon_r=1.5$) with an inner cube
    ($2\,$m, $\epsilon_r=2.0$) and a PEC sphere ($0.5\,$m radius).
    Right, a representative PEC airplane with $7\,$m wingspan and
    body length, and a
  $1.5\,$m tail fin.}
  \label{fig:complex-geometries}
\end{figure}

For objects containing
dielectric surfaces, especially multiple stacked dielectric surfaces, our
approach greatly reduces both memory and runtime (See Table
\ref{tab:performance-comparison}). We compare the
performance of
computing a single monostatic response at normal incidence for both
of our glass
cubes (Figure \ref{fig:correctness-dielectric}) and a nested diamond
and glass cube with PEC sphere in the
center (Figure
\ref{fig:complex-geometries}). All three objects are symmetric with
respect to vertical and horizontal polarization. The nested cube
evaluates performance on
objects with multiple interreflections and complex field paths. 101
frequencies from 1 to 3 GHz were computed. Monte Carlo used 10 strata per
wavelength, with 4 samples in each area, and the standard algorithm used an
equivalent ray density of 20. Both used a fixed max bounce count of
9. Runtimes were
averaged across 10 trials, with 3 warmup runs to allow for Dr. Jit to
cache the GPU kernels.
Peak GPU memory usage was reported by Dr. Jit.

For the deterministic solver, two values are given for
memory. Dr. Jit uses a copy-on-write optimization to allocate
more memory only when a buffer is written. For single region, convex
dielectrics, this can greatly reduce the memory usage as the rays
quickly leave the scene. Such optimizations, however, are difficult in
optimized code not using a higher level library. As such, we force
Dr. Jit to allocate the necessary stack memory at each material
boundary to characterize what a typical SBR implementation, without
this optimization, would use. Both numbers are reported, but
even with this optimization, our method still requires half the memory.

First comparing glass cubes, we see little difference in runtime
because there
are no complex interaction between dielectric surfaces in the object.
The ray state tree
(Figure \ref{fig:dielectric_tree}) for these objects is
degenerate, as rays leaving
do not reflect back. In both methods, the number of rays traced is
linear with the path
length; however, even in this simple case, we see large improvements
in the memory usage.
The standard algorithm cannot, in general, tell that this degenerate
tree structure
will form, so it must save the state of the rays at each hit and
trace both the reflected
and transmitted rays. This leads to a memory growth exponential in
the number of paths.
By using less memory, larger scenes can fit on a GPU and longer, more
complex multipath
effects can be calculated. The nested cube also sees a similar order
of magnitude improvement
in memory usage, but also realizes a 4 times speed up in runtime. For
the nested cube, the
state tree looks more similar to the tree shown in Figure
\ref{fig:dielectric_tree}. The
Monte Carlo algorithm does not need to trace each individual path
like the standard algorithm
needs to, resulting in a faster solver. Using techniques like Fresnel
and Russian roulette sampling,
as explored in the next section, the Monte Carlo algorithm is able to
reduce the necessary computation
even further.

\begin{figure}[thp]
  \begin{subfigure}{1.0\linewidth}
    \centering
    \includegraphics[width=0.8\linewidth]{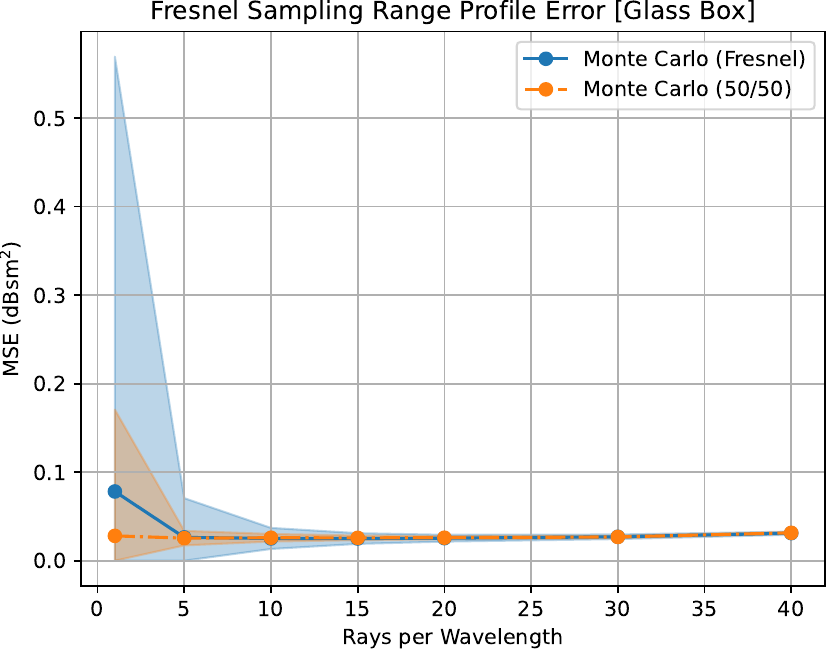}
  \end{subfigure}

  \medskip
  \begin{subfigure}{1.0\linewidth}
    \centering
    \includegraphics[width=0.8\linewidth]{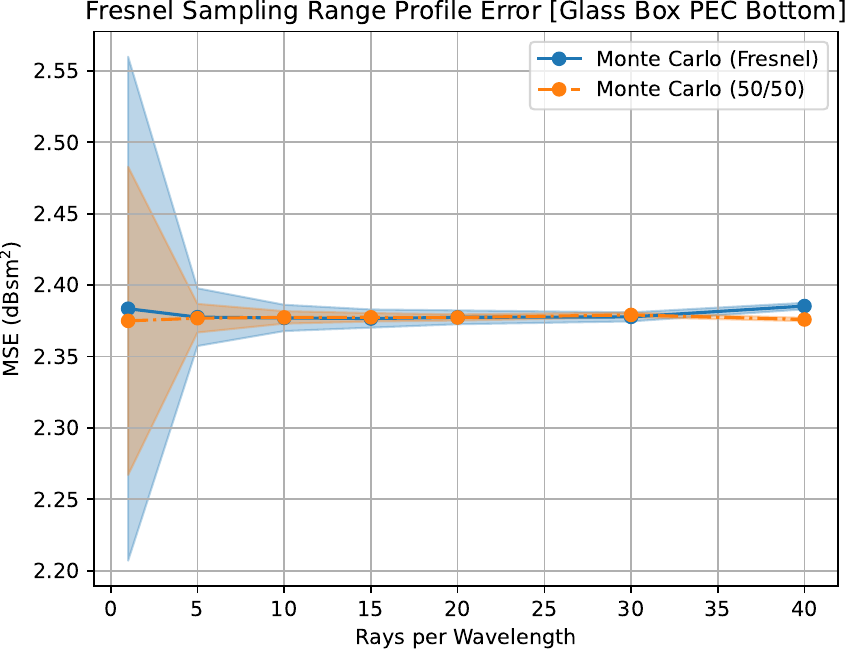}
  \end{subfigure}

  \medskip
  \begin{subfigure}{1.0\linewidth}
    \centering
    \includegraphics[width=0.8\linewidth]{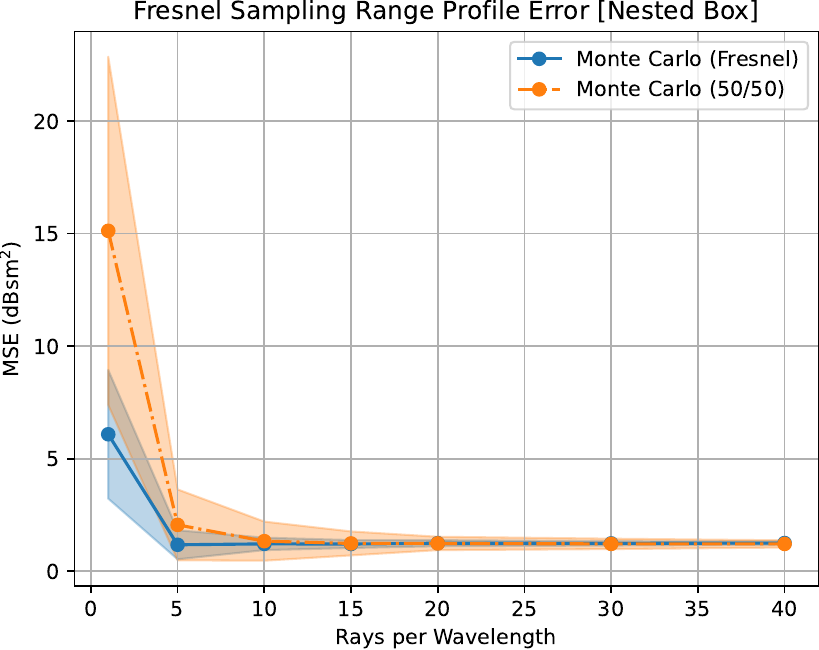}
  \end{subfigure}
  \caption{Fresnel sampling performance on three cube configurations:
    dielectric, PEC-backed dielectric, and nested dielectric. Errors
    are averaged over 100 seeds, with mean and first standard deviation
    shown. Solid lines/areas denote Fresnel sampling; dashed/hatched
    regions indicate 50/50 splitting. Fresnel sampling underperforms
    for single dielectrics due to high transmission losses, but excels
    in nested dielectrics by emphasizing significant paths. This
    suggests Fresnel sampling remains effective for complex, realistic
  objects with multiple dielectric regions.}
  \label{fig:fresnel-sampling-performance}
\end{figure}

\subsection{Sampling}

\subsubsection{Fresnel Sampling}
\label{sec:fresnel-sampling}

We propose using a ray split probability based on the average
reflection coefficient calculated by the Fresnel equation instead of
a simple 50/50 split. We compare performance on the two glass cubes
(Figure \ref{fig:correctness-dielectric}) and nested
dielectric geometry (Figure \ref{fig:complex-geometries}). For the
glass cubes, we compute the error
relative to the analytical solution, while for the nested cubes, we
compare against the standard
algorithm with a ray density of 40 rays per wavelength. We again
average across 100 seed values and
plot both the mean and first standard deviation. The solid line and
area are the Fresnel sampling
results, and the dashed line and hatched area are the 50/50 sampling
results. Results are plotted in
Figure \ref{fig:fresnel-sampling-performance}.

For the glass cube, we find that the Fresnel sampling actually performs worse
than the 50/50 sampling with higher variance, especially at lower
ray densities.
This difference is due to the fact that the reflection coefficient is on the
order of 0.101 and most of the wave's energy leaves the cube on every bounce.
We, however, need to compute long interreflections so need to trace many
internally reflecting rays. The PEC backed cube performs better because the
metal plate forces these reflections. Even with the help, the Fresnel sampling
performs worse than the 50/50 sampling. Fresnel sampling begins to improve
performance on the nested cubes. Because rays do not quickly leave the
scene, the sampling follows scattering rays that quickly leave the
inner layers
and scatter on the outer surface. Though Fresnel sampling does not improve
performance in all scenarios, many objects will have multiple
dielectric regions
where the technique performs best.

\begin{figure}[thp]
  \begin{subfigure}{1.0\linewidth}
    \centering
    \includegraphics[width=0.85\linewidth]{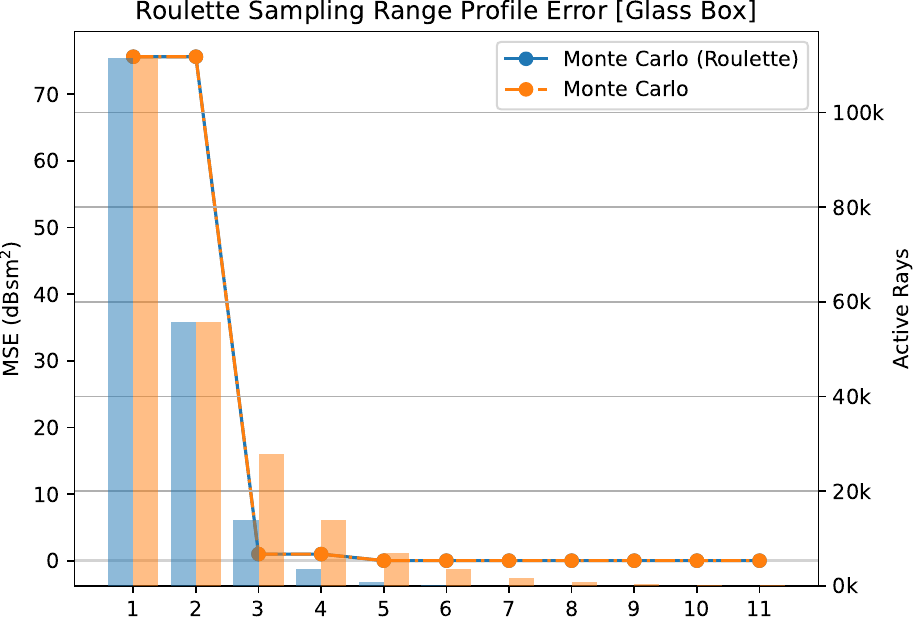}
  \end{subfigure}

  \medskip
  \begin{subfigure}{1.0\linewidth}
    \centering
    \includegraphics[width=0.85\linewidth]{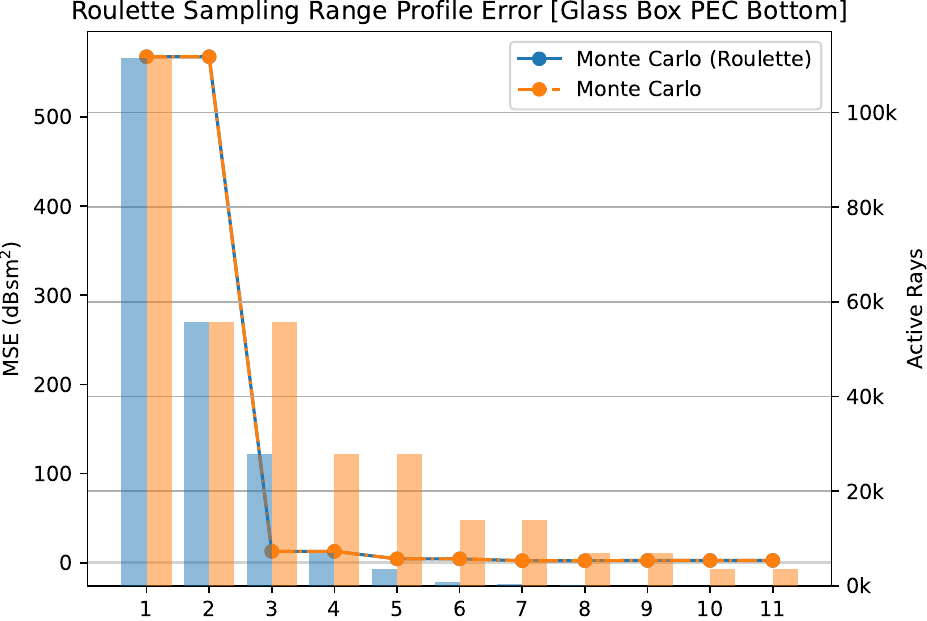}
  \end{subfigure}

  \medskip
  \begin{subfigure}{1.0\linewidth}
    \centering
    \includegraphics[width=0.85\linewidth]{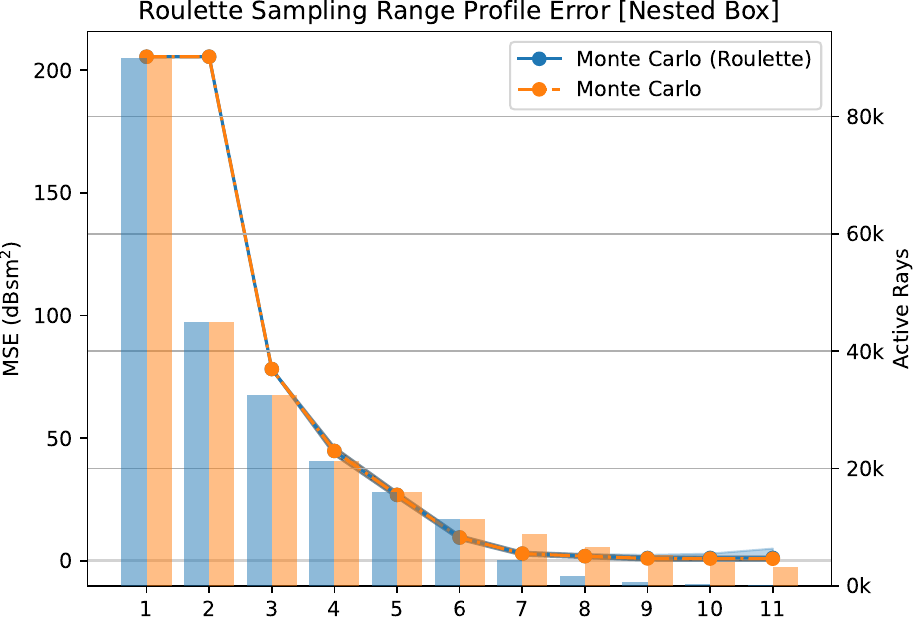}
  \end{subfigure}
  \caption{Russian roulette sampling rapidly reduces active rays,
    halving them after the first bounce compared to full Monte Carlo.
    For single dielectric cubes, accuracy is maintained; for nested
    cubes, variance increases but remains controllable and often
  justified by the performance gain.}
  \label{fig:roulette-sampling-performance}
\end{figure}

\subsubsection{Russian Roulette Sampling}
\label{sec:russian-roulette-sampling}

We also propose using Russian roulette sampling to truncate long paths by
probabilistically killing rays. These long paths arise from large
interreflections common in dielectric materials, though certain PEC geometries
can produce many bounces as well. After a minimum number of bounces,
2 for the glass and pec backed cubes and 6 for the nested cubes, we begin
sampling from a uniform distribution to test if the ray should be
killed based on a fixed probability of 0.5. Though this sampling
technique introduces
slight variance in the estimator, we expect to see an improvement in speed as
long paths are removed early.

Figure \ref{fig:roulette-sampling-performance} represents the active
ray count by the bar graph and the error by the line graphs. This figure
compares the number of active rays at any given $l$ in the path space
with the corresponding error. We find that roulette sampling quickly
decreases the number of rays
after the initial cutoff. Within three bounces, our solver is able to
stop while the deterministic method continues tracing rays until an
overall maximum is reached. For the single
dielectric blocks, we see no difference between the computed scattered
fields and very
little variance in the solution. The ray killing in the nested
dielectrics introduces
no bias for longer paths, but a slight increase in variance is
found as expected. Russian roulette sampling
along with the constant memory usage allows our method to easily
compute long paths
with limited computational overhead.

\begin{figure}[t]
  \centering
  \begin{subfigure}{1.0\linewidth}
    \centering
    \includegraphics[width=1.0\linewidth]{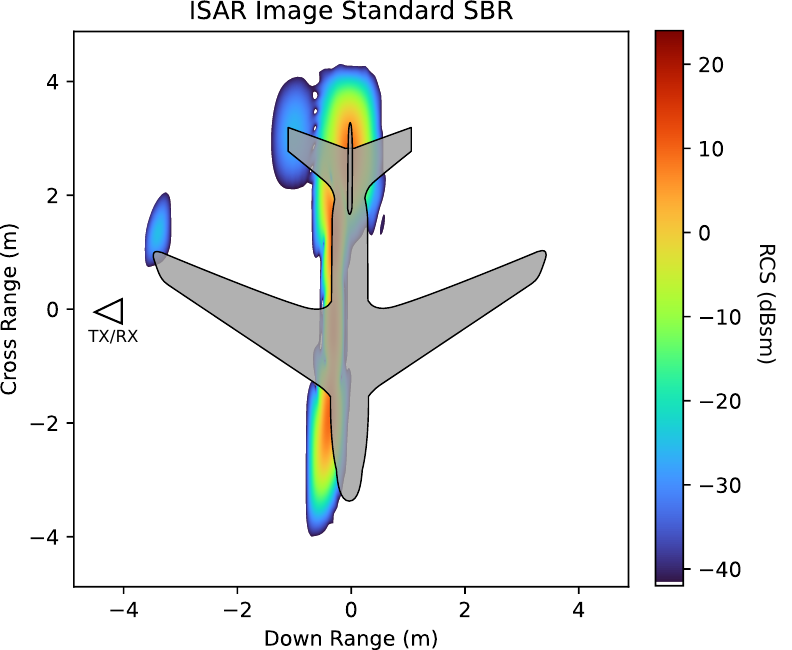}
  \end{subfigure}
  \par\medskip
  \begin{subfigure}{1.0\linewidth}
    \centering
    \includegraphics[width=1.0\linewidth]{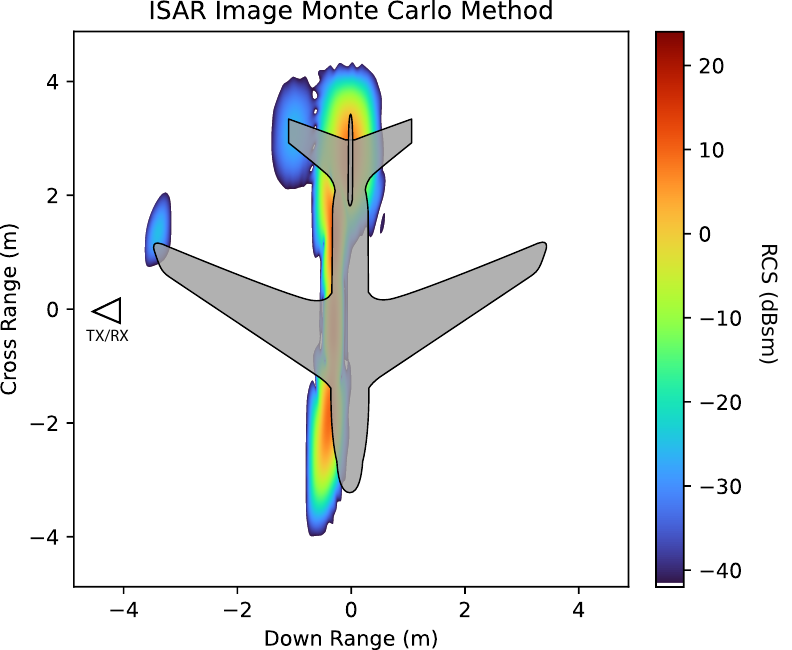}
  \end{subfigure}
  \caption{Comparison of inverse synthetic aperture RADAR images for
    a PEC airplane using standard and Monte Carlo algorithms. Both
    methods accurately capture wing tip scattering and specular
    reflections, with no additional noise observed above the $-40\;dB$
    threshold. Results demonstrate the suitability of the Monte Carlo
  method for input to signal processing applications.}
  \label{fig:isar-images}
\end{figure}

\subsection{Sensitivity to Noise}

\begin{figure}[thp]
  \centering
  \begin{subfigure}{1.0\linewidth}
    \centering
    \includegraphics[width=1.0\linewidth]{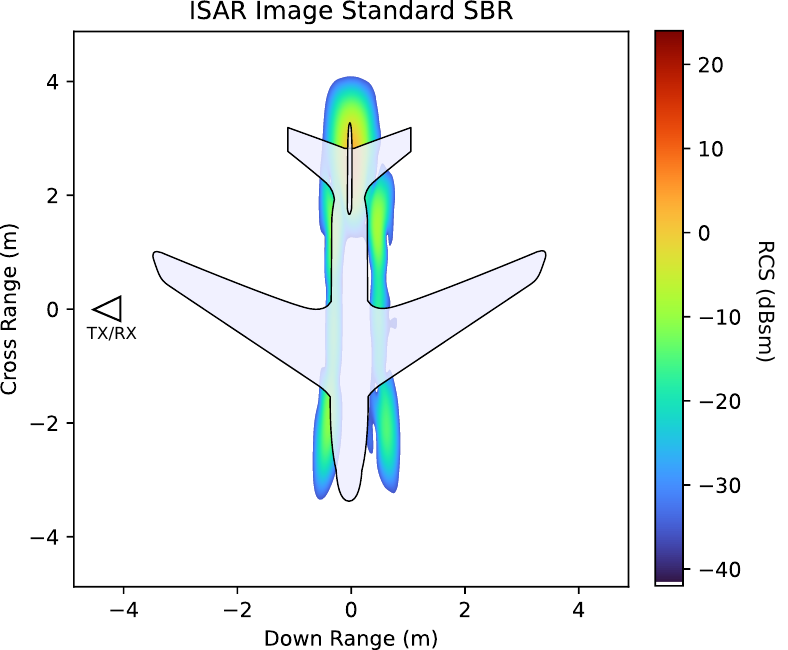}
  \end{subfigure}
  \par\medskip
  \begin{subfigure}{1.0\linewidth}
    \centering
    \includegraphics[width=1.0\linewidth]{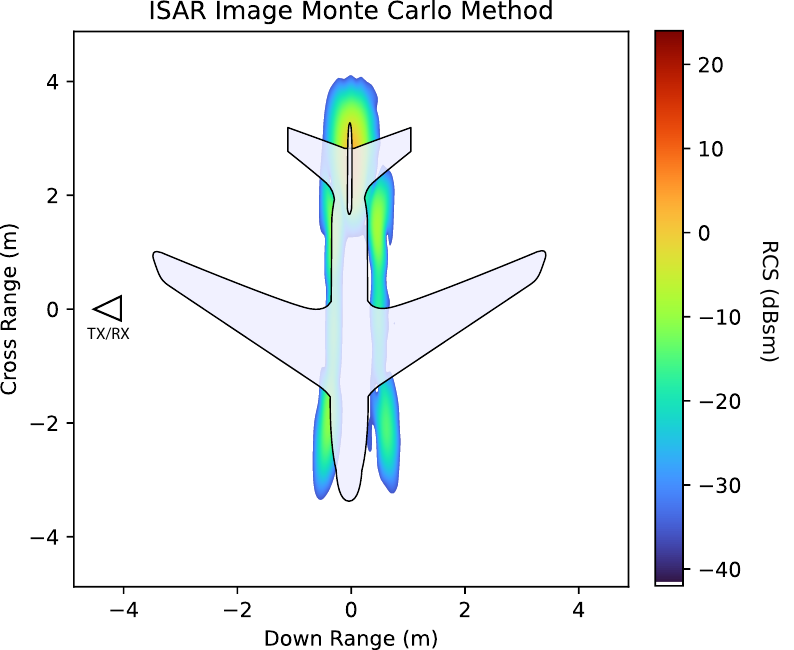}
  \end{subfigure}
  \caption{Comparison of inverse synthetic aperture RADAR images for
    a dielectric airplane ($\epsilon_r=1.5$) using standard and Monte
    Carlo algorithms. Both methods accurately capture specular
    reflections and wall scattering, with no additional noise above
    $-40\;dB$ in the Monte Carlo results. This demonstrates the
    suitability of the Monte Carlo method for downstream signal
  processing applications.}
  \label{fig:isar-images-glass}
\end{figure}

Monte Carlo integration makes higher dimensional integrals easier
to compute at
the cost of additional noise in the answer. Though real world systems expect
noise in their result, if the noise produced by the numerical approximation is
too high, then downstream signal processing can be corrupted. With sufficient
sampling, we are able to get nulls at comparable levels to the classic SBR
algorithm (Figures \ref{fig:correctness-dielectric} and
\ref{fig:range-profile-mci-vs-std}) in range profiles. To
further demonstrate the applicability of our results to down stream tasks, we
compare inverse synthetic aperture (ISAR) reconstructions
\cite{chen_inverse_2014} of the
airplane model (Figure \ref{fig:complex-geometries}) to show that the noise in
our results does not affect the reconstruction. ISARs are incredibly sensitive
to errors in phase, which will cause misplaced scatters which are no longer
grounded to the underlying geometry. Figures \ref{fig:isar-images}
and \ref{fig:isar-images-glass} show that our
method and the standard algorithm have similar performance with an
effective ray density of 20. To match spacing in cross and down
range, 130 frequencies from 1 to 3 GHz along 51 azimuthal angles from
80$^\circ$ to 100$^\circ$. 3 bounces were used on the PEC geometry,
while the max bounce was set to 5 for the dielectric plane. On PEC,
we see that both methods produce ISARs that highlight
the scattering off the wing tips and specular scattering off the fuselage and
vertical stabilizer. No additional noise is found above the $-40 \;
dB$ level. Similar performance is found on the glass airplane
geometry. We see that the specular along with the opposite side
scattering are captured by both techniques. For the PEC case, the runtime
was roughly 5 minutes for both methods, while the dielectric plane
took around 20 minutes in each.
These results demonstrate that our method is suitable for
traditional downstream
tasks for scattered field simulations without onerous sampling
requirements that would increase runtime over the deterministic algorithm.

Our results demonstrate that the Monte Carlo integration technique
brings strong improvements to the SBR algorithm for complex
geometries and difficult to calculate dielectric stack ups. Though
performance is no better for simple shapes, our method uses an order
of magnitude less memory and can be 4 times faster on objects
containing multiple dielectric surfaces. We demonstrate additional
performance improvements using common techniques from Monte Carlo
based path tracing, but by no means have exhausted the extensive path tracing
literature. Our results open up new domains for the SBR algorithm to
compute scenes that are still too expensive for full wave solvers but
require more fidelity than Geometric Optics provides on its own.

\section{Conclusion}

Our work extends the classic SBR algorithm to use Monte Carlo
integration instead of poorly scaling quadrature rules. This change
enables the SBR algorithm to compute scenes faster by decreasing ray
trace calls and better aligning the algorithm with the underlying
parallel hardware. Our results demonstrate this performance gain on
dielectric blocks and nested material objects. The method does have
limitations, and for simpler geometries, there is little benefit as
runtimes are similar and noise is introduced. We also did not
discuss inclusion of additional phenomenology like edge diffraction,
cavity responses, or creeping waves. These are all deficiencies of
the standard SBR algorithm but can be included in our
method using existing techniques and superposition. In future work, higher order
diffraction effects could even be easier to implement using our
method. Another potential avenue is
improved variance reduction techniques. We explored basic
methods like stratified and Fresnel based importance sampling, but
there is a long literature on Monte Carlo integration that could be
leveraged. Notable methods are specular manifold importance sampling
and control variates. Our method improves the
scalability of the SBR algorithm while also enabling exciting avenues
for future work leveraging techniques common in computer graphics.


%

\appendices

\section*{Acknowledgment}

The authors would like to thank Shayne Gerber, Dr. Leo Tchorowski, and
Markus Ferrell for their feedback and comments on early drafts.

\ifCLASSOPTIONcaptionsoff
\newpage
\fi



\bibliographystyle{IEEEtran}
\bibliography{bibtex/bib/references}
%



%

\begin{IEEEbiographynophoto}{Samuel Audia}
  Samuel Audia is a PhD student at the University of Maryland,
  College Park in the departmente of Computer Science. He previously
  received BSc's in Computer Science and Mechanical Engineering from
  the University of Maryland, College Park. Sam also has a MSc from
  Johns Hopkins University in Applied and Computational Mathematics.
  His research focuses on improving computational electromagnetic
  simulations using methods from machine learning and computer graphics.
\end{IEEEbiographynophoto}

\begin{IEEEbiographynophoto}{Dinesh Manocha}
  Dinesh Manocha is a Distinguished University Professor at the
  University of Maryland and Paul Chrisman Iribe Professor of
  Computer Science and Electrical and Computer Engineering. He is
  also the Phi Delta Theta/Matthew Mason Distinguished Professor
  Emeritus of Computer Science at Chapel Hill University of North
  Carolina. Mancha’s research focuses on AI, robotics, computer
  graphics, augmented/virtual reality, and scientific computing, and
  has published more than 730 papers (H-index 135). He has supervised
  46 PhD dissertations, and his group has won 21 best paper awards at
  leading conferences. His group has developed many widely used
  software systems (with 500K+ downloads) and licensed them to more
  than 60 commercial vendors. He is an inventor of 16 patents,
  several of which have been licensed to industry. A Fellow of AAAI,
  AAAS, ACM, IEEE and Sloan Foundation, Manocha is a ACM SIGGRAPH
  Academy Class member and Bézier Award recipient from Solid Modeling
  Association. He received the Distinguished Alumni Award from IIT
  Delhi and the Distinguished Career in Computer Science Award from
  Washington Academy of Sciences. He was also the co-founder of
  Impulsonic, a developer of physics-based audio simulation
  technologies, which Valve Inc. acquired in November 2016. He is
  also a co-founder of Inception Robotics, Inc.
\end{IEEEbiographynophoto}


\begin{IEEEbiographynophoto}{Matthias Zwicker}
  Matthias Zwicker is the Elizabeth Iribe Chair for Innovation and
  the Phillip H. and Catherine C. Horvitz Professor of Computer
  Science, and currently serves as the Chair of the Department of
  Computer Science at UMD. He joined UMD in March 2017 as the
  Reginald Allan Hahne Endowed E-Nnovate Professor in Computer
  Science. He obtained his PhD from ETH in Zurich, Switzerland, in
  2003. From 2003 to 2006 he was a post-doctoral associate with the
  computer graphics group at the Massachusetts Institute of
  Technology, and then held a position as an Assistant Professor at
  the University of California in San Diego from 2006 to 2008. From
  2008-2017, he was a professor in Computer Science at the University
  of Bern, Switzerland, where he served as the head of the Computer
  Graphics Group and as director of undergraduate and graduate
  studies. His research focus is on the intersection of artificial
  intelligence and computer graphics, with the goal of enabling next
  generation AR/VR and computer graphics applications. He has served
  as a papers co-chair and conference chair of the IEEE/Eurographics
  Symposium on Point-Based Graphics, and as a papers co-chair for
  Eurographics and the Eurographics Symposium on Rendering. He has
  been a member of program committees for various conferences
  including ACM SIGGRAPH, ACM SIGGRAPH Asia, and Eurographics, and he
  has served as an associate editor for journals such as the Computer
  Graphics Forum, IEEE TVCG, and the Visual Computer. As an active
  member of the UMD community, he has served as a University Senator
  for the CS department, as a faculty co-lead in developing the new
  Immersive Media Design major, and as a faculty mentor in the
  Gemstone Honors Program.
\end{IEEEbiographynophoto}

\vfill



\end{document}